\documentclass[]{aastex631}

\usepackage{amsmath} 
\usepackage{bm} 
\usepackage{booktabs}
\usepackage{textcomp}
\usepackage{multirow}
\usepackage{graphicx}

\begin{document}

\title{Impact of Large-Scale Structure along Line-of-Sight on Time-Delay Cosmography}

\author[0000-0003-1859-2780]{Shijie Lin}
\affiliation{Institute for Frontier in Astronomy and Astrophysics, Beijing Normal University, Beijing, 102206, China}
\affiliation{School of Physics and Astronomy, Beijing Normal University, Beijing 100875, China}

\author{Bin Hu}
\affiliation{Institute for Frontier in Astronomy and Astrophysics, Beijing Normal University, Beijing, 102206, China}
\affiliation{School of Physics and Astronomy, Beijing Normal University, Beijing 100875, China}

\author{Chengliang Wei}
\affiliation{Purple Mountain Observatory, Chinese Academy of Sciences, Nanjing, 210008, China}

\author{Guoliang Li}
\affiliation{Purple Mountain Observatory, Chinese Academy of Sciences, Nanjing, 210008, China}

\author{Yiping Shu}
\affiliation{Purple Mountain Observatory, Chinese Academy of Sciences, Nanjing, 210008, China}

\author{Xinzhong Er}
\affiliation{Tianjin Astrophysics Center, Tianjin Normal University, Tianjin, 300387, China}

\author{Zuhui Fan}
\affiliation{South-Western Institute for Astronomy Research, Yunnan University, Kunming, 650500, China}

\begin{abstract}

Time-delay cosmography offers a promising and independent method for measuring cosmological distances by monitoring multiple images of gravitationally lensed sources. However, beyond the main deflector, large-scale structure along the line-of-sight (LoS) also deflects travelling light rays via weak lensing (WL). Due to resolution limitations, accurately measuring WL on arcsecond scales remains highly challenging.
In this work, we evaluate the LoS effects on both lensing images and time-delay measurements using a more straightforward, high-resolution N-body simulation that provides a more realistic matter distribution compared to the traditional, computationally cheaper halo rendering method. 
We employ the multi-plane ray tracing technique, which is traditionally utilised to compute WL effects at the arcminute scale, extending its application to the strong lensing regime at the arcsecond scale.
We focus on quadruple-image systems and present the following findings: (1) In addition to a constant external convergence, large-scale structures within a region approximately 2 arcminutes in angular size act as external perturbers, inducing inhomogeneous fluctuations on the arcsecond scale. 
(2) Standard single-plane models suffer from severe geometric degeneracies between foreground shear and main lens ellipticity. By ignoring complex multi-plane couplings, these models fail to accurately recover the external convergence, producing broad, bimodal errors. (3) Explicitly accounting for the LoS shear via the minimal lens model successfully breaks these degeneracies, yielding robust and tightly constrained reconstructions. (4) Uncorrected foreground convergence introduces a $0.5\sim0.7\%$ systematic bias in time-delay distance estimation.
Our findings highlight that advanced LoS modeling is indispensable for precision time-delay cosmography.

\end{abstract}

\section{Introduction} \label{sec:intro}

The Hubble constant ($H_0$), which quantifies the present global expansion rate of the universe, is the central parameter for various cosmology and astrophysics studies. However, the inferred value of this quantity from high redshift cosmic microwave background (CMB) anisotropy from \texttt{Planck} satellite mission \citep{2020A&A...641A...6P} is significantly different from those obtained by the low redshift Cepheid-SNe Ia distance \citep{2016ApJ...826...56R,2022ApJ...934L...7R}. The former measures the statistically isotropic baryonic acoustic oscillation pattern imprinted in the spatial distribution of the background radiation intensity/polarization at the recombination epoch ($z\simeq1100$). By assuming the standard cosmological model ($\Lambda$CDM) which is currently dominated by the cosmological constant ($\Lambda$) and cold dark matter (CDM) components, we can transfer the measurement of sound horizon at the recombination epoch to the present global expansion rate, and the result indicates $H_0=67.4\pm0.5~{\rm km/s/Mpc}$ \citep{2020A&A...641A...6P}. The latter, which is also dubbed the local distance ladder approach, is a cosmology independent method. It utilises distance indicators based on stellar physics to calibrate the SNe Ia's luminosity roughly within 40 Mpc, such as the Cepheid period-luminosity relation, Tip of Red Giant Branches (TRGB), and others. 
Using the Hubble Space Telescope's (HST) Cepheid, \texttt{SH0ES} team reported in 2016 $H_0=73.24\pm1.74~{\rm km/s/Mpc}$~\citep{2016ApJ...826...56R}, and an updated value of  $H_0=72.51\pm1.54~{\rm km/s/Mpc}$ in 2022 \citep{2022ApJ...934L...7R}. 
The 2022 estimate was anchored to NGC 4258, which is a point of consensus between the \texttt{SH0ES} and Carnegie-Chicago Hubble Program (\texttt{CCHP}) teams.
The \texttt{CCHP} team, using TRGB stars as their primary calibrators, found $H_0=69.8\pm 1.9~{\rm km/s/Mpc}$ using HST data \citep{2019ApJ...882...34F}, and an improved value of $H_0=70.4\pm 1.9~{\rm km/s/Mpc}$ by combining HST and James Webb Space Telescope (JWST) data \citep{2025ApJ...985..203F}. 
Notably, while the TRGB-based results are gradually converging toward those derived from Cepheids, heated debates remain over several challenging systematics, such as crowding and sample selection effects \citep{2024ApJ...977..120R,2025ApJ...985..203F}. This highlights the necessity of an independent third distance measurement.

There are many other cosmological distance measurement methods \citep{2019NatAs...3..891V}, including cosmic chronometers \citep{2002ApJ...573...37J,2019JCAP...03..043J} and gravitational wave standard sirens \citep{1986Natur.323..310S,2017Natur.551...85A}. Among these ``third-party'' methods, time-delay cosmography (TDCosmo) stands out as one of the most promising approaches \citep{1964MNRAS.128..307R,2016A&ARv..24...11T,2020MNRAS.498.1420W,2022A&ARv..30....8T,2024SSRv..220...48B}.
The basic principle behind the TDCosmo method involves monitoring the brightness variations of multiple images in a galaxy-AGN strong lensing (SL) system, where the foreground galaxy acts as a gravitational lens, splitting the light from a single source into several ``virtual'' images. The AGN in the system serves as the source. Due to the thermal instability of the AGN disk, brightness fluctuates by about 0.1 magnitudes over weeks to months. Since the different lensing images follow different light paths, they show the same light curve but with a time delay. This time delay consists of a geometric component and a gravitational (Shapiro) component, with its overall magnitude being inversely proportional to the Hubble constant. By capturing high-resolution images of the lensing system with space telescopes, such as HST and JWST, we can accurately reconstruct the lens mass distribution, which is crucial for isolating the cosmological signal. Over the past decades, the \texttt{COSMOGRAIL} project \citep{2005A&A...436...25E,2020A&A...640A.105M} has monitored the light curves of several strong lensing systems over the course of 1.5 decades. With this data, the \texttt{TDCOSMO} collaboration combined high spatial resolution lensing images from the HST to provide measurements of the Hubble constant. In 2020, \texttt{TDCOSMO} reported $H_0 = 74.5^{+5.6}_{-6.1}~\mathrm{km/s/Mpc}$ based on 7 lenses \citep{2020A&A...643A.165B}, and updated in 2025 $H_0 = 72.1^{+4.0}_{-3.7}~\mathrm{km/s/Mpc}$ with the addition of one more lens \citep{2025arXiv250603023T}. Its mean value aligns with the \texttt{SH0ES}'s Cepheid-SN distance indicator. The next goal of the TDCosmo program is to achieve a $1\%$ precision in the estimation of the Hubble parameter by monitoring approximately 50-ish galaxy-AGN SL systems \citep{2019adds.book..353S}.

To achieve this goal, various systematics in TDCosmo must be thoroughly verified, including dynamical mass estimation, microlensing contamination debiasing, and others \citep{2020A&A...643A.165B}. Among these, one of the most critical systematics is the Mass Sheet Degeneracy (MSD), which arises from both the uncertainty in the lens's internal mass profile and the external convergence contributed by large-scale structures along the LoS.\citep{2014MNRAS.439.2432J, 2014MNRAS.443.3631M, 2017ApJ...836..141M, 2017JCAP...04..049B, 2021MNRAS.504.2224L, 2021JCAP...08..024F,2024JCAP...10..055J}. The external convergence suggests that we can introduce a constant surface mass density sheet along the LoS and rescale the position of the source. This transformation leaves the observed lensing image unchanged but alters the time delays between multiple images—directly impacting the inferred value of the Hubble constant. The arcminute scale Weak Lensing (WL) measurement provides a means to break the external MSD. 
Several efforts have been done by combining SL and WL analyses, particularly for galaxy-AGN systems such as Q0957+561 \citep{2009ApJ...697.1793N,2018A&A...614A...8C,2018MNRAS.480.3140W,2018ApJ...867..107W,2019MNRAS.490..613S,2020MNRAS.498.3241B,2018MNRAS.477.5657T,2020MNRAS.498.1406T,2021JCAP...04..010K}.

To gain a deeper understanding of LoS effects, multi-plane ray tracing techniques are essential \citep{2010MNRAS.405.2579O,2012MNRAS.421.2553X,2016ApJ...828...54L,2016MNRAS.462.3255C,2019A&A...624A..54S,2021JCAP...08..024F,2024JCAP...08..021D,2024JCAP...10..055J}. Non-linear effects accumulated along LoS cannot be accurately captured by modeling them solely as tidal perturbations in the primary lens plane. A full three-dimensional treatment is required \citep{2014MNRAS.440..870T,2017ApJ...836..141M}. However, incorporating additional parameters into such 3D models significantly increases computational demands—an important consideration given the expected volume of future data \citep{2010MNRAS.405.2579O,2015ApJ...811...20C}. Another significant challenge lies in the quality of WL signal reconstruction. Current cosmic shear measurements typically achieve galaxy number densities of approximately 10-30 galaxies per square arcminute. At this density, the signal-to-noise ratio (SNR) at arcsecond scales is dominated by shot noise. To address this limitation, a series of simulation-based studies have been conducted. For instance, by matching the observed galaxy number density near the lens, several works have estimated the external convergence using the Millennium Simulation \citep{2009A&A...499...31H,2010ApJ...711..201S,2013ApJ...766...70S,2011MNRAS.410.2167F,2013MNRAS.432..679C,2013ApJ...768...39G,2017MNRAS.467.4220R,2017MNRAS.465.4895W,2024A&A...689A..87W}. However, these approaches are largely statistical in nature and may be insufficiently precise for characterizing the environments of individual lens systems \citep{2011ApJ...726...84W}. To mitigate this limitation, several observationally motivated methods have been proposed. These include rendering galaxy and halo distributions to match deep HST observations of individual TDCosmo lenses, such as RXJ1131 and HE0435, thereby providing a more accurate reconstruction of the local mass distribution \citep{2018MNRAS.477.5657T,2020MNRAS.498.1406T}.

In the traditional ray tracing combined with halo/galaxy rendering approach \citep{2014MNRAS.443.3631M,2017ApJ...836..141M,2017JCAP...04..049B, 2021MNRAS.504.2224L,2025arXiv250604201T}, LoS perturber galaxies are assigned within the light beam based on observational data or mock catalogs. This method does not require high-resolution cosmological simulations across all scales, from the large-scale structure down to the individual lensed galaxy–AGN system. Instead, it generates the intervening structure by sampling dark matter halos from a standard halo mass function and placing them within the simulated light-cone volume. While computationally efficient, this approach randomly populates dark matter halos along the LoS based on mass and redshift distributions, thereby failing to capture the realistic large-scale structure clustering that is inherently present in N-body simulations.

In this paper, we aim to quantify the time-delay error induced by large-scale matter clustering. To achieve this goal, we evaluate the LoS effects on both lensing images and time-delay measurements using a more straightforward, high-resolution N-body simulation \texttt{ELUCID} \citep{2014ApJ...794...94W,2016ApJ...831..164W}, which provides a more realistic matter distribution than the commonly used, computationally cheaper halo-rendering approach. 
Section~\ref{sec:methods} outlines the theoretical lensing formalism, while Section~\ref{sec:mock_data} describes our numerical methodology, including multi-plane ray tracing and mock data generation. Before presenting the statistical analysis, we validate the numerical stability of our approach in Section~\ref{sec:numerical_validation}. Our main results and model reconstructions are presented in Section~\ref{sec:results}. Finally, we summarize our conclusions and discuss several specific implications of the LoS effects in Section~\ref{Sec: Conclusion and discussion}.


\section{Lensing Formalism} \label{sec:methods}

In this section, we first introduce the standard single-deflector lensing framework and extends it to include LoS perturbations. We then present the minimal lens model, an efficient approach that reduces complex multi-plane distortions into a set of effective LoS parameters within the tidal approximation regime.

\begin{figure}[!htp]
\centering
\includegraphics[width=1\textwidth]{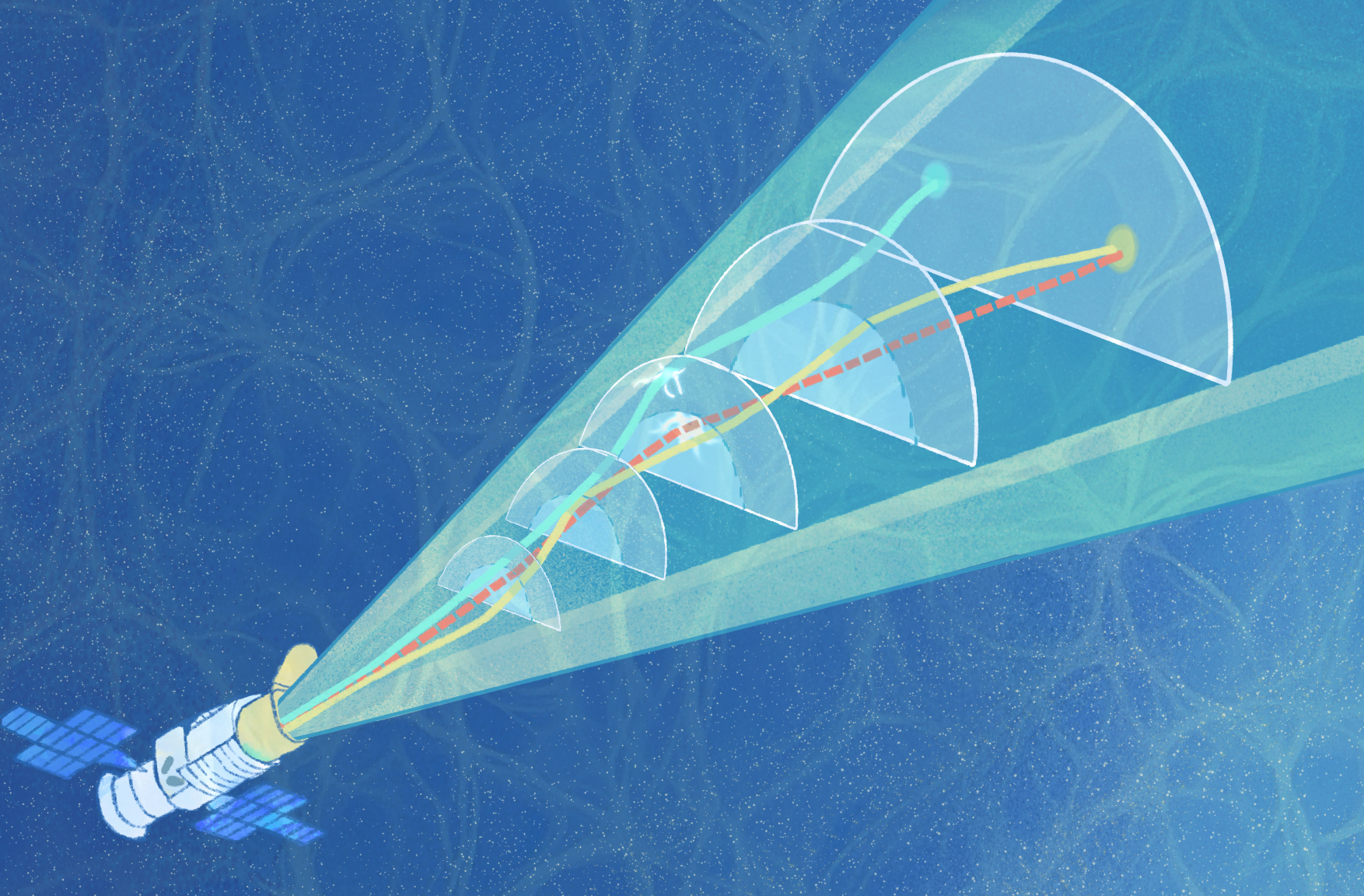}
\caption{A cartoon figures for the physical ray of multi-plane lensing. The red line represents the main lens case, while the yellow line indicates scenarios where the LoS effects are considered within a small region. In the context of large-scale structures, the light ray is deflected by a significant uniform deflection angle as it traverses the universe, following the blue dashed line.
\label{fig:RaysCartoon}}
\end{figure}

\subsection{Lensing Equation for Single Main Lens}
For a single main lens case, the time delay between two images is  
\begin{equation}\label{eq:ddt}
\Delta T_{\rm AB} = \frac{D_{\rm dt}}{c} \Delta\Phi_{\rm AB}, 
\end{equation}
where \( c \) is the speed of light, $\Phi(\boldsymbol{\theta},\boldsymbol{\beta})$ is the Fermat potential, defined as
\begin{equation}\label{eq:ddt2}
\Phi(\boldsymbol{\theta},\boldsymbol{\beta}) = \frac{(\boldsymbol{\theta}-\boldsymbol{\beta})^2}{2}-\psi(\boldsymbol{\theta}).
\end{equation}
Here, \( D_{\rm dt} \) is the time delay distance, which is the combination of relative distances \( (1+z_{\rm d})\frac{D_{\rm l}D_{\rm s}}{D_{\rm ls}} \). \( D_{\rm l} \), \( D_{\rm ls} \), and \( D_{\rm s} \) are the angular diameter distances from the observer to the lens, from the lens to the source, and from the observer to the source, respectively. \( \boldsymbol{\theta} \) and \( \boldsymbol{\beta} \) represent the lens and source plane angular position vector of the light, and
$\psi(\boldsymbol{\theta})$ denotes the lensing potential, which describes how the mass of the lensing object distorts the light rays passing near it and is defined as
\begin{equation}\label{eq:potential}
\psi(\boldsymbol{\theta}) = \frac{1}{\pi} \int \frac{\kappa(\boldsymbol{\theta}')}{|\boldsymbol{\theta} - \boldsymbol{\theta}'|} d^2\boldsymbol{\theta}'\;,
\end{equation}
where \( \kappa(\boldsymbol{\theta}) \) is the dimensionless surface mass density
\begin{equation}\label{eq:suface density}
\kappa(\boldsymbol{\theta}) = \frac{\Sigma(\boldsymbol{\theta})}{\Sigma_{\text{cr}}}\;.
\end{equation}
\( \Sigma(\boldsymbol{\theta}) \) is the projected mass density of the lens, and \( \Sigma_{\text{cr}} \) is the critical surface mass density, given by:
\begin{equation}
\Sigma_{\text{cr}} = \frac{c^2}{4\pi G} \frac{D_{\rm s}}{D_{\rm l} D_{\rm ls}},
\end{equation}
where \( G \) is the gravitational constant.  Given the lensing potential $\psi(\boldsymbol{\theta})$, one can derive the deflection angle of light as it passes near a gravitational lens:
\begin{equation}\label{eq:deflection_angle}
\boldsymbol{\alpha}(\boldsymbol{\theta}) = \boldsymbol{\nabla}\psi(\boldsymbol{\theta})\;,
\end{equation}
and the inverse magnification matrix $\boldsymbol{\mathcal{A}}(\boldsymbol{\theta})$:
\begin{align}\label{eq:inverse_magnification}
\boldsymbol{\mathcal{A}}(\boldsymbol{\theta}) &= \boldsymbol{I} - \boldsymbol{\nabla}\boldsymbol{\alpha}(\boldsymbol{\theta}) \\
    &= \begin{bmatrix} 1-\kappa(\boldsymbol{\theta})-\gamma_1(\boldsymbol{\theta}) & -\gamma_2(\boldsymbol{\theta}) \\
-\gamma_2(\boldsymbol{\theta}) & 1-\kappa(\boldsymbol{\theta})+\gamma_1(\boldsymbol{\theta})
\end{bmatrix}
\end{align}
where $\boldsymbol{I}$ is the $2 \times 2$ identity matrix.

At last, the final lensing equation can be written as
\begin{equation}
    \boldsymbol{\beta}(\boldsymbol{\theta}) = \boldsymbol{\theta}-\boldsymbol{\alpha}(\boldsymbol{\theta})\;.
\end{equation}

\subsection{Lensing Potential Induced by LoS Perturbers on a Single Plane}\label{subs:Lensing Perturbation}
To establish a clear physical picture of the LoS perturbations, it is instructive to first consider the gravitational potential perturbation projected on a single lens plane, denoted as $\psi_p$. One can perform a Taylor series expansion with respect to the coordinates of the lens plane \( (\theta_1,\theta_2) \), expressed as follows \citep{2001astro.ph..2341K}:
\begin{equation}\label{eq: pertrubation}
\psi_p(\boldsymbol{\theta}) \approx \psi_{p,0}(\boldsymbol{0}) + \left .\frac{\partial\psi_p}{\partial\theta_a}\right |_{\boldsymbol{\theta=0}}\theta_a + \left .\frac{1}{2}\frac{\partial^2 \psi_p}{\partial\theta_a\partial\theta_b}\right |_{\boldsymbol{\theta=0}}\theta_a\theta_b + \left .  \frac{1}{3!} \frac{\partial^3 \psi_p}{\partial\theta_a\partial\theta_b\partial\theta_c}\right |_{\boldsymbol{\theta=0}}\theta_a\theta_b\theta_c + \cdots
\end{equation}
where {\it a,b,c} follow Einstein summation convention, implying summation over repeated indices. The zeroth-order term, \(\phi_{p,0}\), represents an unobservable constant offset in the potential and can therefore be omitted.
The linear term, expressed as \( \left .\frac{\partial\psi_p}{\partial\theta_a}\right |_{\boldsymbol{\theta=0}}\theta_a \equiv {\boldsymbol{\alpha}(\boldsymbol{0})\boldsymbol{\theta}} \) , corresponds to a uniform external deflection affecting the entire system, which also remains unobservable in the lensing system, shown in Fig \ref{fig:RaysCartoon} and will be discussed in detail below. Although this linear term modifies the geometrical path and lensing potential of different rays in time delay, it can be shown that these changes cancel out. Thus, the linear term can also be dropped.

The second-order term \(\left .\frac{1}{2}\frac{\partial^2 \psi_p}{\partial\theta_a\partial\theta_b}\right |_{\boldsymbol{\theta=0}}\theta_a\theta_b \) is known as the tidal regime of the strong lensing system. One can introduce the partial shear matrices as the first-order derivative of the deflection angle,
\begin{equation}
\boldsymbol{I}-\boldsymbol{\mathcal{A}}_{\rm ext} = \left. \frac{\partial\boldsymbol{\alpha}} {\partial\boldsymbol{\theta}} \right|_{\boldsymbol{\theta=0}}=
\begin{pmatrix}
\kappa_{\rm ext}+\gamma_{1,\rm ext} & \gamma_{2,\rm ext} \\
\gamma_{2,\rm ext} & \kappa_{\rm ext} -\gamma_{1,\rm ext}
\end{pmatrix},
\label{eq:tidal_expansion}
\end{equation}
where parameters \( \kappa_{\rm ext} \), \( \gamma_{1 ,\rm ext} \) and \( \gamma_{2,\rm ext} \) are referred to as the external convergence, which introduces the MSD effect, and the external shear which stretches the images of the source, transforming circles into ellipses, respectively. 
Beyond the tidal regime, the 3rd order term \(\left .  \frac{1}{3!} \frac{\partial^3 \psi_p}{\partial\theta_a\partial\theta_b\partial\theta_c}\right |_{\boldsymbol{\theta=0}}\theta_a\theta_b\theta_c\), namely the flexion effect, originates from the inhomogeneity of convergence and shear. This term is described by four additional parameters~\citep{2006MNRAS.365..414B,2008A&A...485..363S,2023JCAP...08..065T}
\begin{align}
\bm{\mathcal{F}} &= \begin{pmatrix} \mathcal{F}_1 \\ \mathcal{F}_2 \end{pmatrix} = \begin{pmatrix} \frac{\partial\kappa_{\rm ext}}{\partial\theta_1} \\ \frac{\partial\kappa_{\rm ext}}{\partial\theta_2} \end{pmatrix}, \\
\bm{\mathcal{G}} &= \begin{pmatrix} \mathcal{G}_1 \\ \mathcal{G}_2 \end{pmatrix} = \begin{pmatrix} \frac{\partial\gamma_{1 ,\rm ext}}{\partial\theta_1} - \frac{\partial\gamma_{2,\rm ext}}{\partial\theta_2} \\ \frac{\partial\gamma_{1 ,\rm ext}}{\partial\theta_2} + \frac{\partial\gamma_{2,\rm ext}}{\partial\theta_1} \end{pmatrix},
\end{align}
with the first flexion \(\mathcal{F} \) inducing skewed images and the second flexion \( \mathcal{G} \) producing asymmetrical distortions. Under the influence of flexion, lensed images often appear as arc-like structures. 

\subsection{The LoS Minimal Lens Model}

While Section \ref{subs:Lensing Perturbation} establishes the physical effects of perturbations on a single plane, realistic LoS structures are distributed across multiple planes at various redshifts. In principle, each of these individual planes induces local perturbations that can be described by the expansion introduced above. However, expanding the potential on every plane and assigning independent parameters is impractical. To account for multi-plane tidal perturbations along the LoS without over-parameterizing the model, the lensing system can be reduced to a \textit{minimal lens model} governed by the equation~\citep{1987ApJ...316...52K,2021JCAP...08..024F,2024JCAP...10..055J}:
\begin{equation}
    \tilde{\boldsymbol{\beta}} = \boldsymbol{\mathcal{A}}_{\rm LoS}\boldsymbol{\theta} - \boldsymbol{\nabla}\psi_{\rm eff}(\boldsymbol{\theta})
\label{eq:LoSMinimal}
\end{equation}
where $\tilde{\boldsymbol{\beta}} \equiv \boldsymbol{\mathcal{A}}_{\rm od}\boldsymbol{\mathcal{A}}_{\rm ds}^{-1}\boldsymbol{\beta}$ represents the modified source position , and $\boldsymbol{\mathcal{A}}_{\rm LoS} \equiv \boldsymbol{\mathcal{A}}_{\rm od}\boldsymbol{\mathcal{A}}_{\rm ds}^{-1}\boldsymbol{\mathcal{A}}_{\rm os}$ is the total effective LoS amplification matrix.
These amplification matrices ($\boldsymbol{\mathcal{A}}_{\rm od}$, $\boldsymbol{\mathcal{A}}_{\rm ds}$, and $\boldsymbol{\mathcal{A}}_{\rm os}$) characterize the distortions from foreground, background, and overall LoS perturbers, respectively.
Explicitly, each inverse amplification matrix $\boldsymbol{\mathcal{A}}_{ij}$ (where $ij \in \{\rm od, ds, os\}$) can be decomposed into its corresponding effective convergence $\kappa_{ij}$ and shear components $\gamma_{1,ij}$ and $\gamma_{2,ij}$:
\begin{equation}
    \boldsymbol{\mathcal{A}}_{ij} = \begin{pmatrix} 1 - \kappa_{ij} - \gamma_{1,ij} & -\gamma_{2,ij} \\ -\gamma_{2,ij} & 1 - \kappa_{ij} + \gamma_{1,ij} \end{pmatrix}
\end{equation}
To first order in these weak perturbations, the components of the total effective LoS matrix $\boldsymbol{\mathcal{A}}_{\rm LoS}$ can be linearly approximated as 
\begin{equation}\label{eq: TidalLoS}
    \kappa_{\rm LoS} \approx \kappa_{\rm od} - \kappa_{\rm ds} + \kappa_{\rm os},\ \boldsymbol{\gamma}_{\rm LoS} \approx \boldsymbol{\gamma}_{\rm od} - \boldsymbol{\gamma}_{\rm ds} + \boldsymbol{\gamma}_{\rm os}.
\end{equation}
In Eq.~\eqref{eq:LoSMinimal}, the foreground inverse amplification matrix $\boldsymbol{\mathcal{A}}_{\rm od}$ directly modifies the position at which the main deflector's potential must be evaluated, resulting in the effective lensing potential $\psi_{\rm eff}(\boldsymbol{\theta}) \equiv \psi[\boldsymbol{\mathcal{A}}_{\rm od}\boldsymbol{\theta}]$. To see the precise expressions of the components in these matrices, we refer reader to~\cite{2021JCAP...08..024F} for details.

The inclusion of these multi-plane tidal corrections requires adjusting the predicted arrival time of each images. The relative time delay $t(\theta, \beta)$ experienced by a lensed ray is expressed as: 
\begin{equation}
    t(\boldsymbol{\theta}, \boldsymbol{\beta}) = \frac{D_{\rm dt}}{c} \left[ \frac{1}{2}(\boldsymbol{\theta} - \boldsymbol{\beta^{\prime}}) \cdot \boldsymbol{\mathcal{A}}_{\rm LoS}(\boldsymbol{\theta} - \boldsymbol{\beta^{\prime}}) - \psi_{\rm eff}(\boldsymbol{\theta}) \right]
\end{equation}
where $\boldsymbol{\beta^{\prime}} \equiv \boldsymbol{\mathcal{A}}_{\rm os}^{-1}\boldsymbol{\beta}$.

\section{Mock Data Generation} \label{sec:mock_data}

Having established the analytical framework for LoS perturbations, we now describe the numerical pipeline used to generate our mock catalog of quadruply-lensed AGN systems. To fully capture the non-linear, higher-order effects of a continuous matter distribution, we employ a full multi-plane ray tracing algorithm applied to high-resolution N-body cosmological simulations.
\subsection{Multi-Plane Ray Tracing}\label{subs:raytracing}
When accounting for LoS effects from the simulation data, one can use the multiple-lens-plane formalism \citep{2009A&A...499...31H,2016ApJ...828...54L,2018ApJ...853...25W}. We begin by approximating the convergence of the i-th plane \( \kappa_{i}(\boldsymbol{\theta}_i) \). Under the Born approximation and assuming a source at infinity, this can be expressed as~\citep{2013MNRAS.435..115B}
\begin{equation}\label{eq:kappa_slice}
    \kappa_{i}(\boldsymbol{\theta}_i) = \frac{3 H_0^2 \Omega_{\rm m} }{2c^2} \, (1+z_i)D(\chi_i) \int_{\chi_{i}-\frac{1}{2}\Delta\chi}^{\chi_{i}+\frac{1}{2}\Delta\chi} \mathrm{d}\chi' \, \delta(\chi',\boldsymbol{\theta}_i),
\end{equation}
where \( \boldsymbol{\theta}_i \) denotes the angular position on the \( i \)-th lens plane, \( \Omega_m \) is the matter density parameter, \( D(\chi_i) \) is the comoving angular diameter distance to comoving distance \( \chi_i \), \( \Delta\chi \) is the thickness of each lens plane slice, and \( \delta \) is the matter overdensity within the simulated light-cone volume.

The lensing potential \( \psi_i(\boldsymbol{\theta}_i) \) is related to the convergence via the two-dimensional Poisson equation
\begin{equation}
    \boldsymbol{\nabla}^2\psi_i(\boldsymbol{\theta}_i) = 2\kappa_{i}(\boldsymbol{\theta}_i).
\end{equation}
Ray tracing through multi-plane to the \( i \)-th lens plane is then performed recursively using
\begin{align}
    \boldsymbol{\theta}_k &= \boldsymbol{\theta} - \sum_{i=0}^{k-1} 
    \frac{D(\chi_k - \chi_i)}{D(\chi_k)} \, \boldsymbol{\alpha}_i, \\
    \boldsymbol{\mathcal{A}}_k &= \boldsymbol{I} - \sum_{i=0}^{k-1} 
    \frac{D(\chi_k - \chi_i)}{D(\chi_k)} \, \boldsymbol{U}_i \boldsymbol{\mathcal{A}}_i,
\end{align}
where the deflection angle \( \boldsymbol{\alpha}_i \) is the gradient of the lensing potential, \( \boldsymbol{\alpha}_i = \boldsymbol{\nabla} \psi_i \), and \( \boldsymbol{U}_i = \boldsymbol{\nabla} \boldsymbol{\alpha}_i \) is the distortion matrix, containing the second derivatives of the potential.
Finally, the source angular position \( \boldsymbol{\beta} \) and the inverse magnification matrix \( \boldsymbol{\mathcal{A}} \) are obtained at \( k = N + 1 \), where \( N \) is the total number of lens planes.

The travel time of the light ray is determined by a combination of the geometric configuration and the lensing potential. Between two planes labeled by {\it i} and {\it i}+1, one can derive the time delay relative to the unperturbed straight path \citep{2021MNRAS.504.2224L}:
\begin{equation}
t_{i,i+1} = \frac{1 + z_i}{c} \frac{D_{i} D_{i+1}}{D_{i,i+1}} \left[\frac{\|\boldsymbol{\theta}_i-\boldsymbol{\theta}_{i+1}\|^2}{2}- \frac{D_{i,i+1}D_s}{D_{i+1}D_{is}} \psi_i(\boldsymbol{\theta}_i)\right],
\label{eq:timedelay1}
\end{equation}
where \( z_i \) is the redshift of the {\it i}-th lens plane. The total time delay between two images is the sum of the time delay between each two nearby planes:
\begin{equation}
\Delta T_{\rm AB} = \sum_{i=1}^N t_{i,i+1}^{\rm A} - \sum_{i=1}^N t_{i,i+1}^{\rm B}.
\label{eq:timedelay2}
\end{equation}

\subsection{N-body Simulation and Simulated Light-Cone Volume Construction}\label{subs:light_cone}
\begin{table}[h]
    \centering
    \caption{Parameters of the ELUCID}
    \label{table:ELUCID}
    \begin{tabular}{@{}cc|cc@{}}
        \toprule
        \( \Omega_{\rm m} \) & 0.282  &  \( h \)  & 0.697  \\ 
        \( \Omega_{\Lambda} \) & 0.718  & \( \sigma_{8} \)  & 0.82 \\
        \( \mathrm{m}_{\rm p} / (10^{10} h^{-1} \rm M_{\bigodot}) \)  & 0.034  & \( l_{\rm soft}/h^{-1}\rm kpc \)  & 3.5 \\ \bottomrule
    \end{tabular}
\end{table}
We generate a realistic matter distribution using a cosmological N-body simulation. 
In this work, we use \texttt{ELUCID} simulation \citep{2014ApJ...794...94W}. This simulation runs with \( 3072^3 \) dark matter particles with a box size of 500 \(h^{-1}\)Mpc. The cosmological parameters of the N-body simulation are listed in Table \ref{table:ELUCID}. We divide the simulation boxes into sets of small cubes with 100 \(h^{-1} \)Mpc on each side, and pile them together to cover the entire simulated light-cone volume from redshift 0 to 2.12. 
We construct a large, flat-sky simulated light-cone volume extending to a redshift of 2.12, with an edge length of approximately 13 arcmin and a pixel resolution of 0.05 arcsec. This simulated light-cone volume is divided into 76 slices, each with a comoving thickness of 50 \( h^{-1}\)Mpc. For each slice, we project the dimensionless surface mass density $\kappa(\boldsymbol{\theta})$ to its effective redshift by Eq.~\eqref{eq:kappa_slice}, and compute the corresponding lensing potential and deflection angle using Fast Fourier Transforms on each lens plane.

\subsection{Mocking the Lens Systems}\label{subs:synthetic_obs}
Then we simulate 100 quadruply-lensed AGN TDCosmo systems, which include extended images from host galaxies, point images from AGN, and corresponding time delays. 
To gather sufficient statistical samples for analyzing LoS effect, we fix the comoving distances of the main lens and source at redshifts \( z_{l} \sim 0.78 \) and \( z_{s} \sim 2.12 \), respectively.
Then, we randomly select 100 directions as the centers of each simulated light-cone volume with \( N_{\rm size} = 2048 \) and the pixel size is 0.05 arcsec. The matter distribution of the main lens is assumed to be a Singular Isothermal Ellipsoid (SIE) density profile with the Einstein radius \( \theta_{E} \sim 1.3\) arcsec and random ellipticity. We perform the same ray tracing method as described in the previous section to obtain the light position and the distortion matrix of all lens planes.

Using data on light position and distortion matrix across all lens planes, we generate light distribution of the source galaxy and lens galaxy as an elliptical S\'{e}rsic profile through \texttt{lenstronomy} \citep{2018PDU....22..189B,2021JOSS....6.3283B}\footnote{https://lenstronomy.readthedocs.io/en/latest/}. 
We assume observations with a background noise characterized by a root mean square of 0.016 ${\rm photon}\,s^{-1}~{\rm pixel}^{-1}$ and an exposure time of 500\,s.
A Gaussian point spread function model with a FWHM of 0.135 arcsec is used. The median SNR for extended images, evaluated over 100 lens realizations, is approximately 50.
For point sources, we used the image-finding method described in \cite{2021MNRAS.504.2224L} to locate the AGN component at the center of the source galaxy. A \( 5\% \) error is introduced in the time delay measurements to mimic the real observation. We focus on quadruply-lensed systems and calculate the time delays between each image using Eq.~\eqref{eq:timedelay1} and (\ref{eq:timedelay2}).



\section{Numerical Validation} \label{sec:numerical_validation}

Before presenting the statistical analysis of our simulated time-delay strong lensing systems, it is essential to validate the numerical stability of our multi-plane ray tracing pipeline. In this section, we determine the optimal angular window size required to achieve convergence in the effective lensing potential. These validation steps ensure that our simulated observables are robust from artificial boundary effects.

\begin{figure}
\centering
\includegraphics[width=0.8\textwidth]{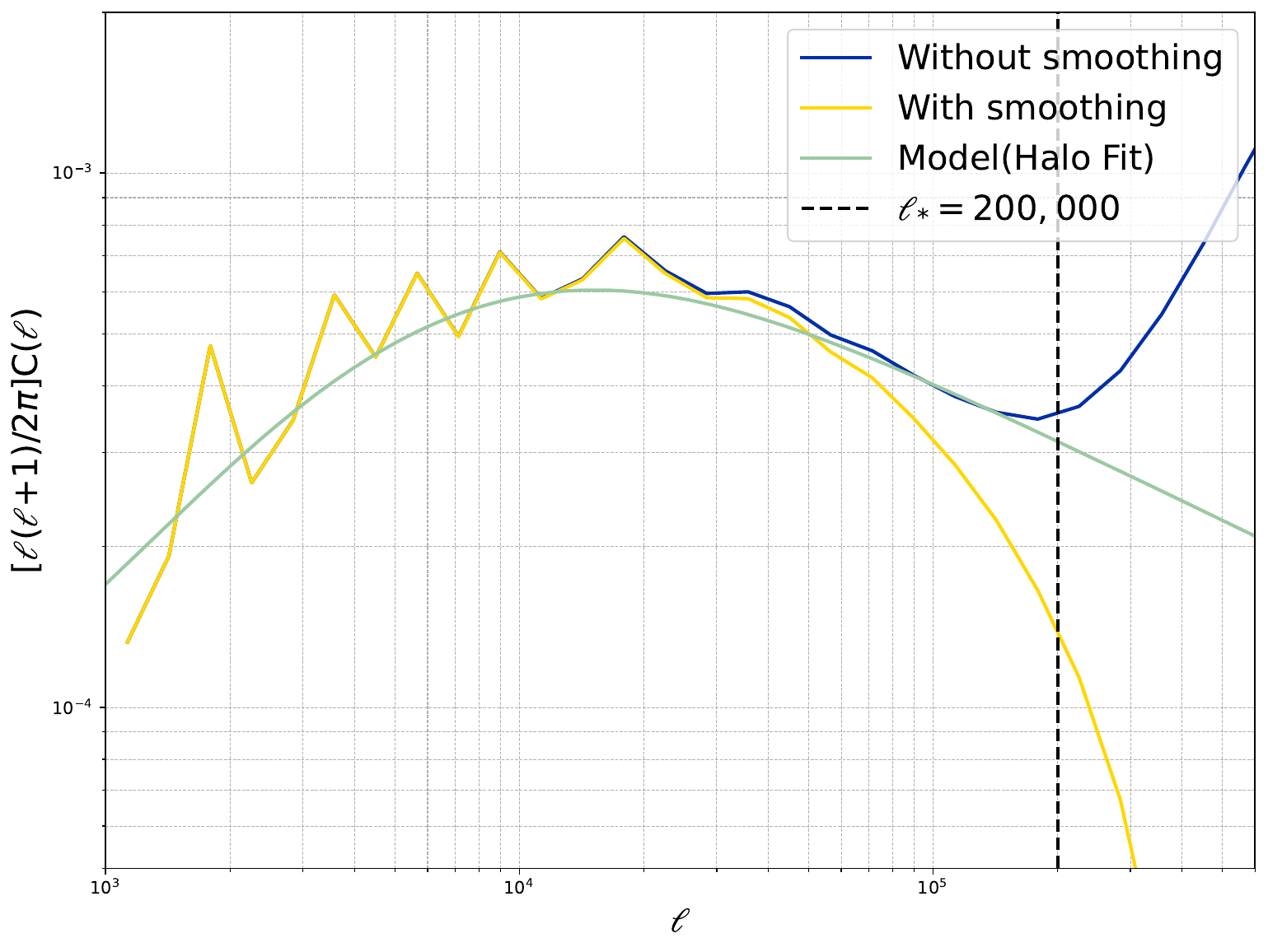}
\caption{The matter–matter angular power spectrum measured from the \texttt{Elucid} simulation.
The blue and yellow curves show the power spectrum without and with Gaussian smoothing, respectively. 
For comparison, the green curve shows the Halofit model prediction \cite{2012ApJ...761..152T}. 
The black dashed vertical line indicates the smoothing scale at $\ell_{*}=2\times10^{5}$.
\label{fig:Cls}}
\end{figure}

\subsection{Smoothing the Shot Noise} \label{subsec:smoothing}
The finite mass resolution of simulation particles introduces shot noise at small scales. To mitigate its impact on our findings, we apply a Gaussian filter with a smoothing scale of $\ell_{*}=2\times10^{5}$, corresponding to an angular size of approximately 3 arcsecs. Fig.~\ref{fig:Cls} shows the matter-matter angular power spectrum measured from \texttt{Elucid} simulation. Before smoothing, the power spectrum shows a significant excess at sub-arcsecond scales, primarily due to shot noise. After applying Gaussian smoothing, the shot noise contribution is substantially suppressed at small scales, ensuring the reliability of our results for structures larger than the smoothing scale. Comparing the green and yellow curves, Gaussian smoothing filters the shot noise but at the cost of suppressing the cosmological signal, reducing its amplitude by a factor of $\sim$2 at a scale of 3 arcsec.

\subsection{The Uniform Deflection}
During our calculation, we find that the deflection angle induced by the LoS large-scale structures reaches approximately the 1 arcmin scale, which is much larger than the SL deflection angle. More importantly, we find that as the angular size of the simulation window increases, the computed potential at the SL position increases monotonically. This occurs because increasing the target window size includes more environmental mass associated with nearby large-scale structures. 

To achieve convergence of the potential value as the angular size increases, we must subtract a uniform deflection angle across the entire angular window. This uniform deflection angle corresponds precisely to the linear term in Eq.~\eqref{eq: pertrubation}, which shifts the geometrical path but has no impact on the relative lensed image positions or the associated time delays. A detailed discussion of this can be found in Section 3.1.3 of~\cite{2021JCAP...08..024F}. A schematic illustration of this procedure is shown in Fig.~\ref{fig:RaysCartoon}. The cyan curve and spot indicate the ``true'' ray-tracing path and source position, respectively. The yellow spot marks the image position as deflected by LoS structures. The red line represents the ray path perturbed solely by the main lens. Within the smaller semi-circular regions, local small-scale structures contribute mainly to higher-order perturbations from nearby matter, causing the rays to follow the path indicated by the yellow line. In contrast, mass located farther from the lens center primarily contributes to the linear perturbation term, producing a nearly uniform deflection across the entire strong lensing system. This effect is illustrated by the cyan line: at each lens plane, the light is uniformly deflected, leading to a shift in the apparent source position relative to the observer’s LoS.

\begin{figure}
\centering
\includegraphics[width=1\textwidth]{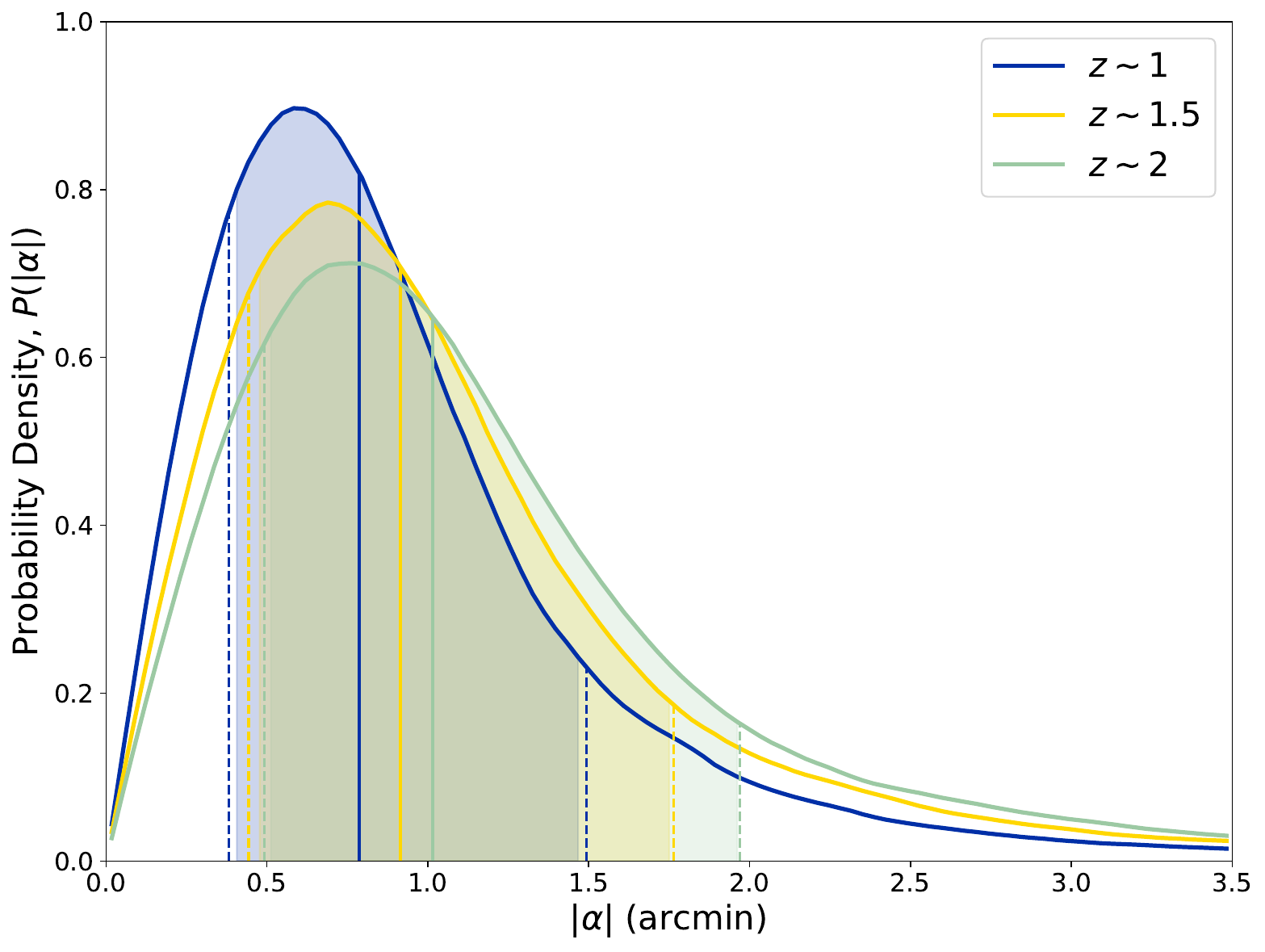}
\caption{Distributions of the deflection angle, $|\alpha|$, for source redshifts $z \sim 1$ (blue), $1.5$ (yellow), and $2$ (green). The solid and dashed vertical lines indicate the median and the 16th and 84th percentiles of the distributions, respectively.
\label{fig:Deflection}}
\end{figure}

To quantify this effect in our simulation, we apply the full-sky ray-tracing technique described in \citet{2018ApJ...853...25W} to compute the deflection angles. Fig.~\ref{fig:Deflection} presents the distribution of the total LoS deflection angles for various source redshifts. For most sources at redshift $z\sim2$, the deflection due to the linear perturbation is about 1 arcmin. This magnitude is consistent with findings in recent work using the MillenniumTNG simulation \citep{2024MNRAS.533.3209F}. 

We present a specific instance in the upper panel of Fig.~\ref{fig:case_potential}, demonstrating that the lensing potential values within the typical Einstein radius do not converge as the size of the calculation region increases. As discussed, this linear term has no observable effects on the strong lensing observables. To mitigate the numerical errors arising from this large linear component, we numerically subtract the averaged constant deflection angle within the window from each lens plane. After implementing this correction, the reconstructed lensing potential exhibits excellent convergence. This improvement is illustrated qualitatively in the lower panels of the middle and right columns of Fig.~\ref{fig:case_potential}, and is further validated statistically by the distribution over 200 realizations presented in Fig.~\ref{fig:statistics phi}.

\subsection{Relevant Angular Scales for LoS Perturbations} \label{subsec1}
To ensure computational efficiency without sacrificing accuracy, a key methodological question is determining the necessary foreground angular extent required to properly capture the LoS lensing observables. 
Using the \texttt{ELUCID} simulated light-cone volume constructed in Section \ref{sec:mock_data}, we randomly select 200 directional simulated light-cone volumes. For each cone, we extract sub-regions with four different angular edge lengths: 25.6, 51.2, 102.4, and 204.8 arcsec (Size I - Size IV). To evaluate the sensitivity to these scales, we compute the total effective LoS potential projected onto the main lens plane by summing the contributions from all 75 multi-lens planes. The effective cumulative projected potential is given by:
\begin{equation}\label{eq:eff_potential}
    \psi_{\rm LoS}(\boldsymbol{\theta}) = \sum_{i \ne {\rm d}} \frac{D_{i\mathrm{s}}}{D_{\mathrm{s}}} \psi_{i}(\boldsymbol{\theta}_i),
\end{equation}
where $D_{i\mathrm{s}}/D_{\mathrm{s}}$ is the geometric scaling factor that projects each individual lensing potential onto the actual source plane. In our actual numerical implementation, these perturbed coordinates are computed using the multi-plane ray-tracing algorithm, tracking the exact non-linear deflection of the light ray as it propagates through each lens plane; thus, the coordinates $\boldsymbol{\theta}_i$ are evaluated along the perturbed ray path. A concise derivation of Eq.~(\ref{eq:eff_potential}) is provided in Appendix~\ref{app:LoS_potential}.

It is worth noting that projecting all LoS contributions onto the primary lens plane is mathematically a single-plane approximation, which deviates from the rigorous multi-plane assumptions of the LoS minimal model. However, this projection is utilized here as a statistical diagnostic tool rather than for precise parameter inference. Although $\psi_{\rm LoS}$ is not a directly measurable quantity, it serves as the fundamental underlying scalar field for all relevant strong lensing observables. Our primary goal in this section is to statistically determine a safe minimum angular radius where outer WL contributions become negligible. Therefore, we track the overall sensitivity of this effective projected potential. This method provides a proxy that is both computationally efficient and physically robust.

\begin{figure}[!htp]
\centering
\includegraphics[width=1\textwidth]{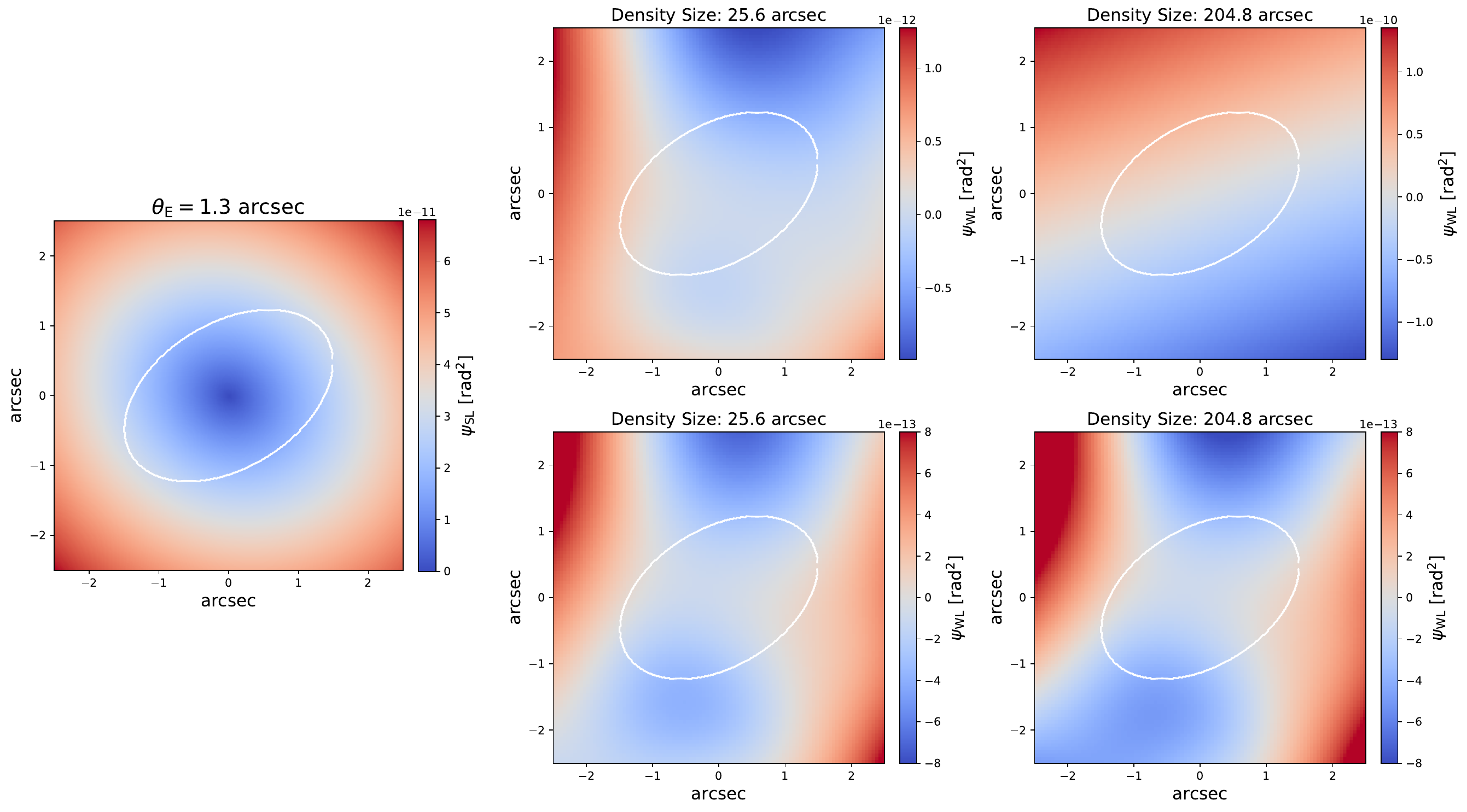}
\caption{Comparison of the primary strong lensing potential and the LoS lensing potential contributions. Left column: The lensing potential of the primary lens ($\psi_{\mathrm{SL}}$) with an Einstein radius of $\theta_{\mathrm{E}} = 1.3\ \mathrm{arcsec}$. Middle and right columns: The LoS lensing potential ($\psi_{\mathrm{WL}}$) calculated from two different density map sizes, $25.6\ \mathrm{arcsec}$ (middle) and $204.8\ \mathrm{arcsec}$ (right) for a single random realization. The upper panels display the raw, un-subtracted LoS potentials, where the potential from the average deflection angle dominates at larger window sizes. The lower panels show the remaining potential perturbations after subtracting the linear components corresponding to the average deflection angles. White contours in all panels delineate the critical curves of the primary system. Note that all panels uniformly display a fixed central FOV of $5\ \mathrm{arcsec} \times 5\ \mathrm{arcsec}$ and the \textit{Density Size} labeled above the columns indicates the spatial extent of the density field used for the integration rather than the plot boundaries.
\label{fig:case_potential}}
\end{figure}

\begin{figure}[!htp]
\centering
\includegraphics[width=1\textwidth]{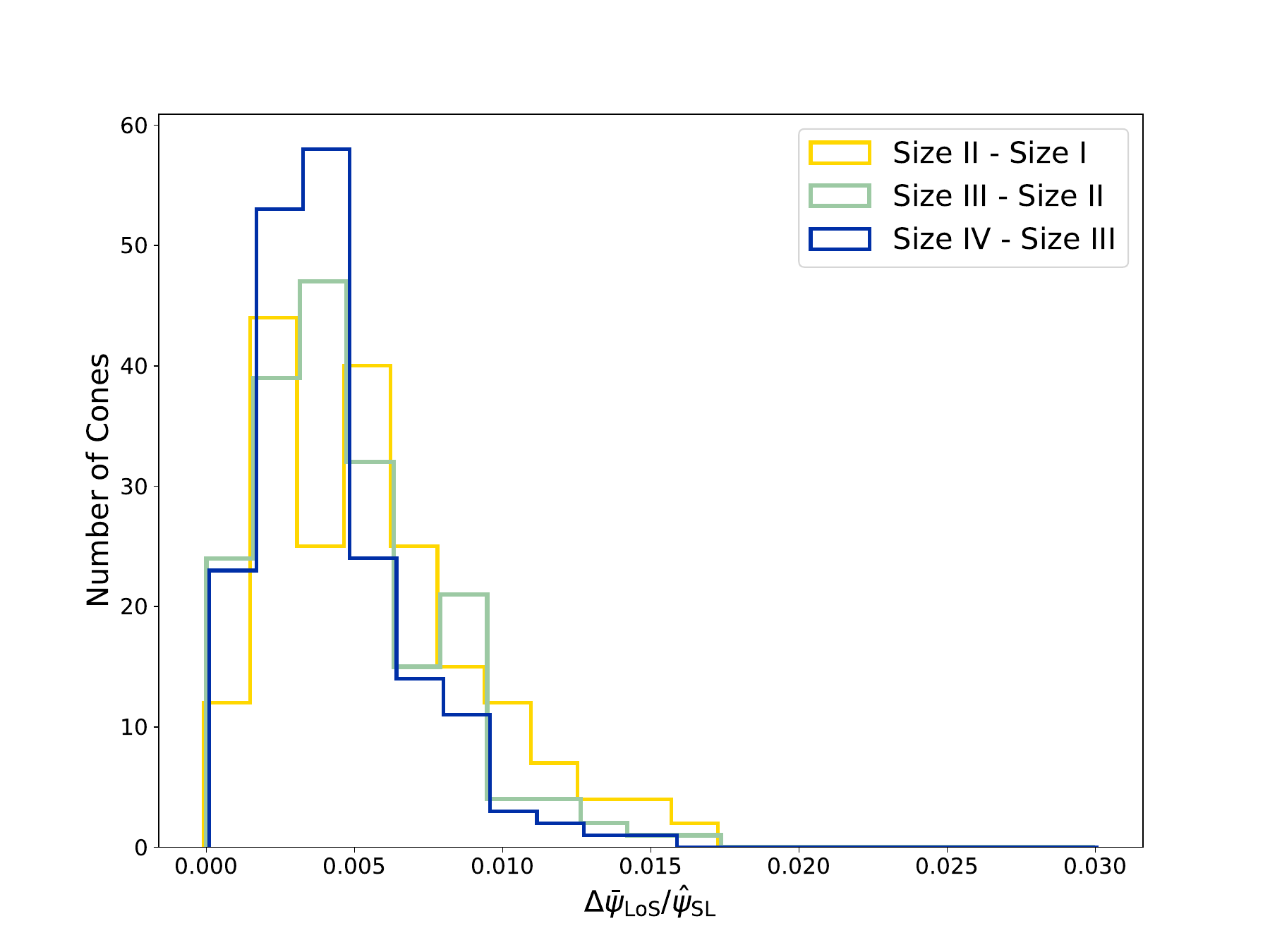}
\caption{Distribution of the differences in the normalized mean effective projected lensing potential, $\Delta\bar{\psi}_{\rm LoS}/\hat{\psi}_{\rm SL}$, evaluated across 200 random light-cone realizations. Here, the normalization factor $\hat{\psi}_{\rm SL}$ is set to $3\times10^{-11}\ \mathrm{rad}^2$, which corresponds to the typical strong lensing potential around the Einstein radius (see the first column of Fig.~\ref{fig:case_potential}). The histograms illustrate the variations in the potential within the central region as the calculation window size is consecutively doubled. Specifically, the yellow, green, and blue stepped lines represent the differences between consecutive sizes: Size II $-$ Size I ($51.2''$ vs. $25.6''$), Size III $-$ Size II ($102.4''$ vs. $51.2''$), and Size IV $-$ Size III ($204.8''$ vs. $102.4''$).
\label{fig:statistics phi}}
\end{figure}

We calculate the mean value \( \bar\psi_{\rm LoS} \) of the this effective lensing potential within the Einstein radius \( \theta_{E} \sim 1.3\) arcsec and show the normalized differences of  as the window size increases in Fig.~\ref{fig:statistics phi}. The typical normalization factor \( \hat\psi_{\rm SL} \) is chosen as $3\times10^{-11} \rm \; rad^{2}$, representing the typical order of magnitude for the lensing potential at the Einstein radius, which estimated from the first column of Fig.~\ref{fig:case_potential}. As shown in Fig.~\ref{fig:statistics phi}, when the calculation window size is increased from 102.4 to 204.8 arcsec, the relative residuals across the 200 realizations converge to within $1\%$. Consequently, a simulated light-cone volume size of 102.4 arcsec is sufficient to capture the relevant LoS effects. We adopt this dimension for the generation of all mock TDSL systems in our subsequent analysis.

\begin{figure}[!htp]
\centering
\includegraphics[width=1\textwidth]{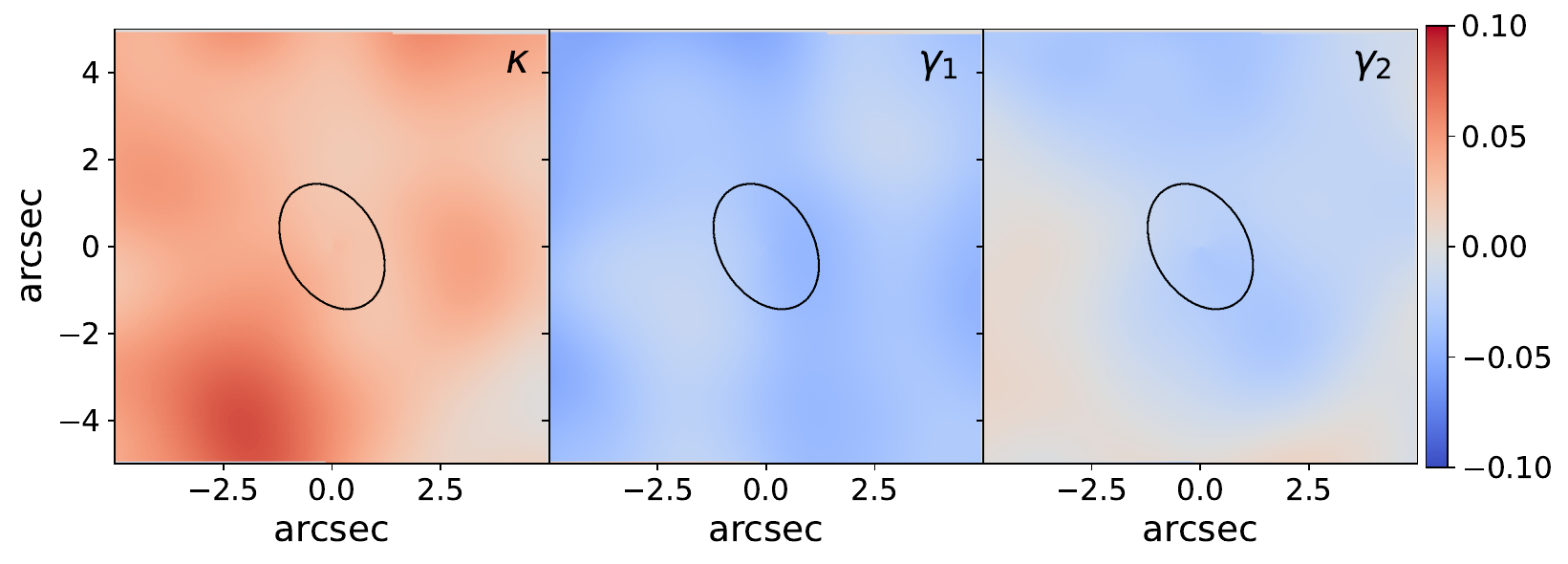}
\caption{LoS convergence and shear maps projected on the main lens plane for a single sample case from the mock datasets. The left, middle, and right panels illustrate the LoS fields $\kappa$, $\gamma_1$, and $\gamma_2$, respectively, centered in the primary lensing region. The central black curve denotes the critical curve of the main deflector.
\label{fig:case_los}}
\end{figure}

With the simulation box size fixed at 102.4 arcsec, we can now robustly assess the WL contamination affecting the strong lens system. As shown in Fig.~\ref{fig:case_los}, the intervening structures introduce higher-order fluctuations into the central lensing potential projected onto the main lens plane. In this case, the fluctuations in convergence and shear are at the percent level, making them comparable in magnitude to the mean constant convergence and shear. We also present the statistical distribution of the standard deviations for these two fields in Appendix~\ref{append:std}. 
Furthermore, we must recall that the density field derived from the \texttt{ELUCID} simulation uses a smoothing scale roughly two to three times larger than the strong lens's typical Einstein radius (see Section~\ref{subsec:smoothing}). Because this smoothing inherently erases the structures at smaller scale, the actual convergence and shear fields in the Universe are likely far more inhomogeneous than those shown in Fig.~\ref{fig:case_los}.
Consequently, the assumption of constant convergence and shear breaks down at the image positions, suggesting that the extra spatial gradients in the shear field strictly require the explicit modeling of single-plane flexion, or even higher-order contributions \citep{2006MNRAS.365..414B, 2008A&A...485..363S}.

\section{Results} \label{sec:results}

Having established the numerical stability of our mock generation pipeline, we now turn to the statistical analysis of the simulated quadruply-imaged systems. 
We model the images and time delay measurements with \texttt{lenstronomy}. The extended light of the lens and source galaxies is modeled with elliptical Se\'{r}sic profiles, while point sources are handled via the \texttt{LENSED\_POSITION} solver in \texttt{lenstronomy} to decouple their influence from the extended emission.

To thoroughly investigate how LoS effects couple with the macro model, we consider two distinct scenarios for the main lens mass profile while leaving the time delay distance ($D_{\rm dt}$) as a free parameter. In the first case, we fix all the main lens mass parameters ($\theta_{\rm E}$, $e_1$, and $e_2$ of the SIE profile) to their true values, ensuring that the optimization is driven entirely by the LoS perturbations. In the second case, we only fix the Einstein radius ($\theta_{\rm E}$) to its true value, leaving the other mass parameters free to vary. In real observations, the inferred Einstein radius is inherently coupled with the LoS convergence due to the MSD. However, in our mock analysis, we have access to the unperturbed $\theta_{\rm E}$. Because our mock lenses are generated using a SIE profile, fixing this true $\theta_{\rm E}$ forces the model to determinate the scaling factor $(1-\kappa_{\rm LoS})/(1-\kappa_{\rm od})$, as formulated in Appendix \ref{sec:appendix_kinematics}.

For each of these two main lens setups, we evaluate our reconstructions across four distinct LoS modeling configurations. These are derived from two base parameterization approaches, both with and without flexion corrections. The first base approach models the perturbations simply as an external convergence (\( \kappa_{\rm ext} \)) and a two-component external shear (\( \gamma_{1,\rm ext}, \gamma_{2,\rm ext} \)) projected onto a single lens plane. The second approach directly fits the explicit multi-plane LoS parameters formulated in the minimal lens model, which include the effective convergences and shears for the LoS structures (\( \kappa_{\rm los}, \gamma_{1,\rm los}, \gamma_{2,\rm los} \)), and the foreground terms (\( \kappa_{\rm od}, \gamma_{1,\rm od}, \gamma_{2,\rm od} \)). This latter model is already implemented via the \texttt{LineOfSight} package in \texttt{lenstronomy}\footnote{https://github.com/lenstronomy/lenstronomy/tree/main/lenstronomy/LensModel/LineOfSight}. To further assess the impact of higher-order spatial derivatives, each of these two base LoS models is evaluated twice: once restricted to the tidal (shear) regime, and once incorporating the single-plane flexion parameters (\( \mathcal{F}_1, \mathcal{F}_2, \mathcal{G}_1, \mathcal{G}_2 \)) which are handled via \texttt{FLEXION} module in \texttt{lenstronomy}\footnote{We defer a full implementation of multi-plane LoS flexion to future work. In this study, we incorporate single-plane flexion to capture higher-order perturbative corrections. This approach aligns with  the treatment of main-plane flexion in TDCosmo \citep{2023JCAP...08..065T} and is conceptually analogous to the effective $\mathcal{F}_{\rm LoS}$ framework discussed by \citet{2024JCAP...08..021D}.}.

\subsection{The Impact from the Foreground Shear}\label{subsec:Foreground Shear}

\begin{figure}
\centering
\includegraphics[width=1\textwidth]{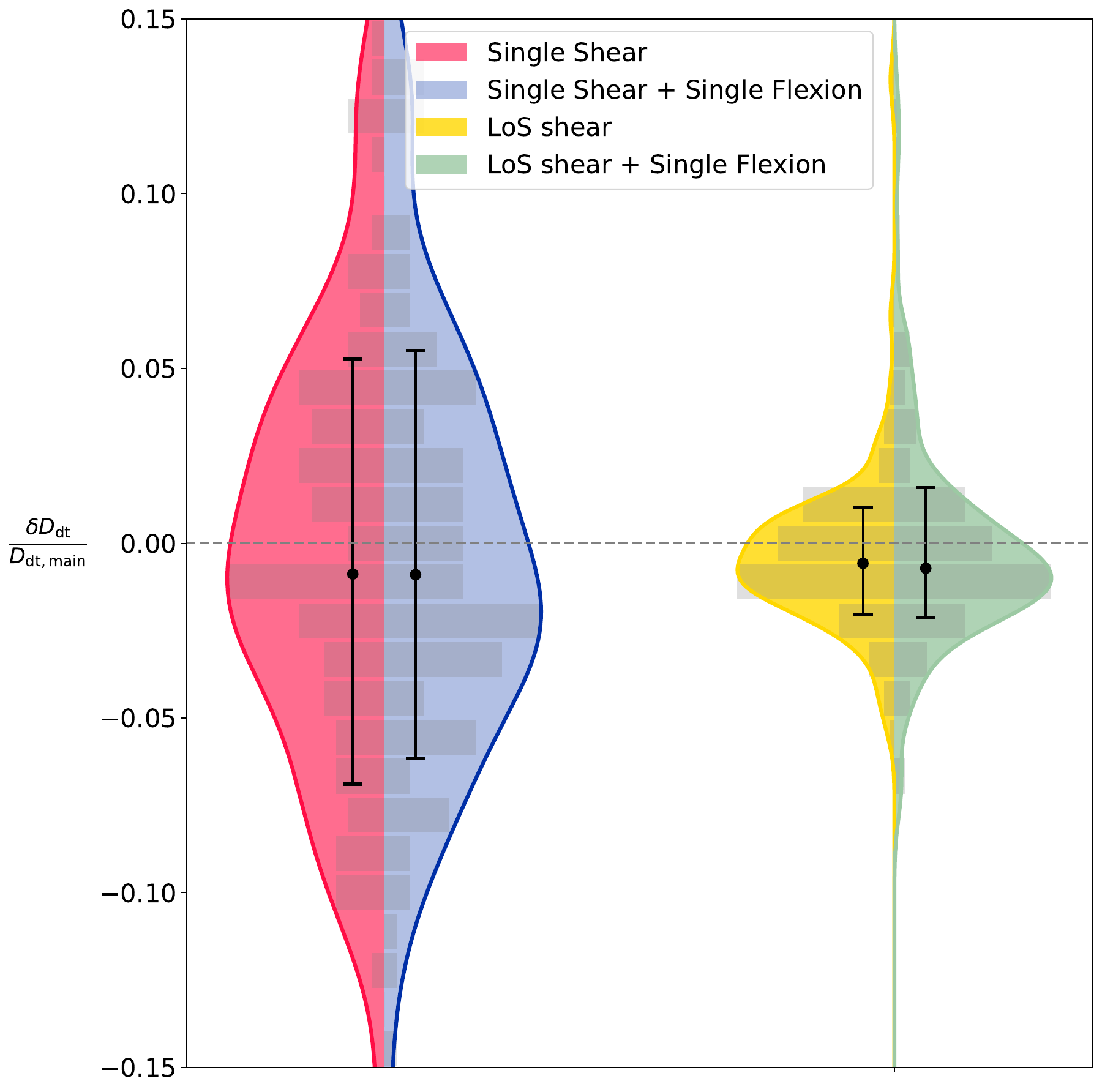}
\caption{Distribution of the modeling bias on the time-delay distance \( D_{\rm dt} \) for 100 mock realizations, with all main lens mass parameters fixed. Split violin plots show the distribution of model-predicted biases for models using only shear (the left half) versus shear plus flexion (the right half). The left and right violins compare the single shear and LoS shear models, respectively. The outer contours represent the kernel density estimates, overlaying the background histograms of the underlying distributions. The central black error bars indicate the median (dot) and the 1$\sigma$ range (horizontal caps). A gray horizontal line at $y = 0$ indicates zero bias.
\label{fig:IntrinsicErrorAllFix}}
\end{figure}

Figs.~\ref{fig:IntrinsicErrorAllFix} and \ref{fig:IntrinsicErrorFixThetaE} present the statistical distributions of $D_{\rm dt}$ modeling bias across 100 mock realizations. Note that $D_{\mathrm{dt}}^{\mathrm{fit}}$, along with all other fitted quantities presented hereafter, corresponds to the median of the posterior distribution reconstructed via Markov Chain Monte Carlo (MCMC) sampling. Fig.~\ref{fig:IntrinsicErrorAllFix} illustrates the $D_{\rm dt}$ bias under idealized conditions, where all main lens mass parameters are fixed to their true values. In contrast, the first and third columns of Fig.~\ref{fig:IntrinsicErrorFixThetaE} demonstrate the impact of relaxing these constraints by keeping only the Einstein radius $\theta_{\rm E}$ fixed and allowing the lens ellipticity parameters to vary. In the relevant panels of both figures, the left violin shows the $D_{\rm dt}$ bias recovered using a standard single-plane external shear model, split to display results without (left half) and with (right half) single-plane external flexion. Similarly, the right violin presents the results for the LoS shear minimal model, comparing inferences with and without single-plane flexion.

It is clearly evident that when all the main lens mass parameters are strictly fixed to their true values, the single-plane perturbation model yields a noticeably worse reconstruction of $D_{\rm dt}$ compared to the explicit LoS modeling. Specifically, the median and 1$\sigma$ bias for the single-plane approach ($-0.009^{+0.053}_{-0.069}$ without flexion, and $-0.009^{+0.055}_{-0.061}$ with flexion) exhibit a significantly wider spread than those of the explicit LoS approach ($-0.006^{+0.010}_{-0.020}$ without flexion, and $-0.007^{+0.016}_{-0.021}$ with flexion). This degraded performance primarily originates from a strong fundamental degeneracy between the main lens ellipticity parameters ($e_1$ and $e_2$) and the foreground shear $\boldsymbol{\gamma}_{\rm od}$. As compared in the first and third columns of Fig.~\ref{fig:IntrinsicErrorFixThetaE}, when these constraints are relaxed and $e_1$ and $e_2$ are left as free parameters in the fit, the single-plane model leads to a seemingly improved reconstruction of the time-delay distance.

\begin{figure}
\centering
\includegraphics[width=1\textwidth]{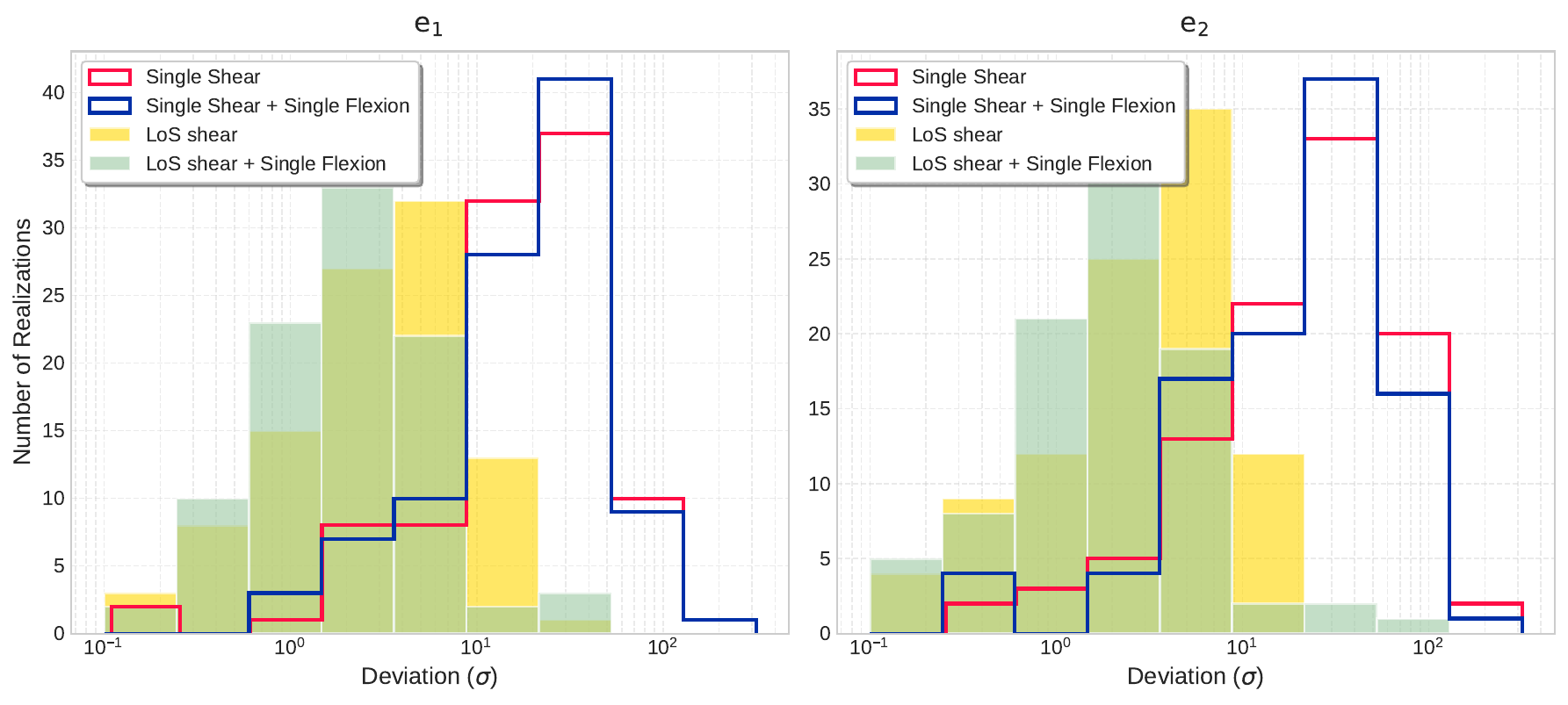}
\caption{Distribution of the deviations between the best-fit and true values for the ellipticity parameters $e_1$ and $e_2$, with only the Einstein radius $\theta_{\rm E}$ fixed The logarithmic x-axis represents the magnitude of the deviation, while the y-axis shows the number of mock systems. Histograms compare the performance of four models: Single Shear (red step line), Single Shear + Single Flexion (blue step line), LoS shear (yellow shaded region), and LoS shear + Single Flexion (green shaded region). Note that both of the Single Shear model exhibits significant bias in ellipticity reconstruction.
\label{fig:e1e2_sigma_deviation}}
\end{figure}

The underlying mechanism driving this increased uncertainty is explicitly revealed when examining the reconstruction of the main lens mass profile. Fig.~\ref{fig:e1e2_sigma_deviation} presents the distributions of the deviations (in terms of $\sigma$) between the best-fit and true values for the lens ellipticity parameters $e_1$ and $e_2$. We observe a severe degeneracy between the foreground shear and the main lens ellipticity. When standard single-plane shear models (Single Shear and Single Shear + Single Flexion, denoted by the red and blue step lines, respectively) are applied to systems with complex LoS perturbations, they fail to recover the true lens ellipticity. For these models, the deviations exhibit a significant long-tail distribution, peaking predominantly between $10\sigma$ and $100\sigma$, while with the LoS modeling (LoS Shear and LoS Shear + Single
Flexion, denoted by the yellow and green shaded region, respectively), the deviations are mostly lower than $10\sigma$.

This severe ellipticity-shear degeneracy strongly aligns with the analytical conclusions drawn by \cite{2021JCAP...08..024F} in section 3.2. Physically, this indicates that the standard single-plane shear model artificially absorbs the unmodeled multi-plane LoS shear perturbations by systematically stretching the main lens ellipticity. In contrast, models that explicitly incorporate the LoS shear, represented by the yellow and green shaded regions, successfully break this degeneracy. By accurately accounting for the foreground perturbations, these models reliably reconstruct the lens profile, dramatically reducing the bias in the recovered ellipticity parameters.

\begin{table*}[htbp]
\centering
\renewcommand{\arraystretch}{1.3} 
\caption{The mean and median bias evaluations for various modeling groups. The left section details the mean estimation bias, standard deviation, t-statistic, and confidence levels ($1-p$) derived from Student's t-tests. The right section provides the median bias, Wilcoxon (W) statistics, Z-scores, and corresponding confidence levels.}
\label{table:stasbias}
\begin{tabular}{l | c c | c c}
\hline
\multirow{2}{*}{\textbf{Group}} & \multicolumn{2}{c|}{\textbf{Mean (t-test)}} & \multicolumn{2}{c}{\textbf{Median (Wilcoxon)}} \\
 & \textbf{Est Bias $\pm$ std} & \textbf{t-stat / (1-p)} & \textbf{Est Bias (W-stat)} & \textbf{Z / (1-p)} \\
\hline
Single Shear                                                      & -0.0002 $\pm$ 0.0425 & -0.038 / 0.0300 & -0.0075 (1792) & -2.52$\sigma$ / 0.9883 \\
Single Shear + Flexion                                            & -0.0028 $\pm$ 0.0301 & -0.914 / 0.6371 & -0.0062 (1798) & -2.50$\sigma$ / 0.9876 \\
Single Shear (with $\kappa_{\rm od}$ Correction)                  &  0.0042 $\pm$ 0.0429 &  0.964 / 0.6625 &  0.0008 (2580) &  0.19$\sigma$ / 0.1531 \\
Single Shear + Single Flexion (with $\kappa_{\rm od}$ Correction) &  0.0016 $\pm$ 0.0304 &  0.508 / 0.3876 & -0.0017 (2336) & -0.65$\sigma$ / 0.4845 \\
LoS Shear                                                         & -0.0010 $\pm$ 0.0391 & -0.265 / 0.2083 & -0.0058 (1865) & -2.27$\sigma$ / 0.9768 \\
LoS Shear + Flexion                                               & -0.0021 $\pm$ 0.0308 & -0.682 / 0.5030 & -0.0072 (1885) & -2.20$\sigma$ / 0.9724 \\
LoS Shear (with $\kappa_{\rm od}$ Correction)                     &  0.0033 $\pm$ 0.0396 &  0.825 / 0.5886 &  0.0005 (2528) &  0.01$\sigma$ / 0.0042 \\
LoS Shear + Single Flexion (with $\kappa_{\rm od}$ Correction)    &  0.0022 $\pm$ 0.0311 &  0.708 / 0.5194 & -0.0011 (2455) & -0.24$\sigma$ / 0.1863 \\
\hline
\end{tabular}
\end{table*}

\subsection{Correcting the MSD effect}

Building on the ellipticity-shear degeneracy discussed previously, we now turn our attention to the MSD and the accurate recovery of the external convergence $\kappa$. Because $D_{\rm dt}$ is inversely proportional to $1 - \kappa$, an unbiased reconstruction of the convergence from LoS perturbers is an absolute prerequisite for precision cosmography.

\begin{figure}
\centering
\includegraphics[width=1\textwidth]{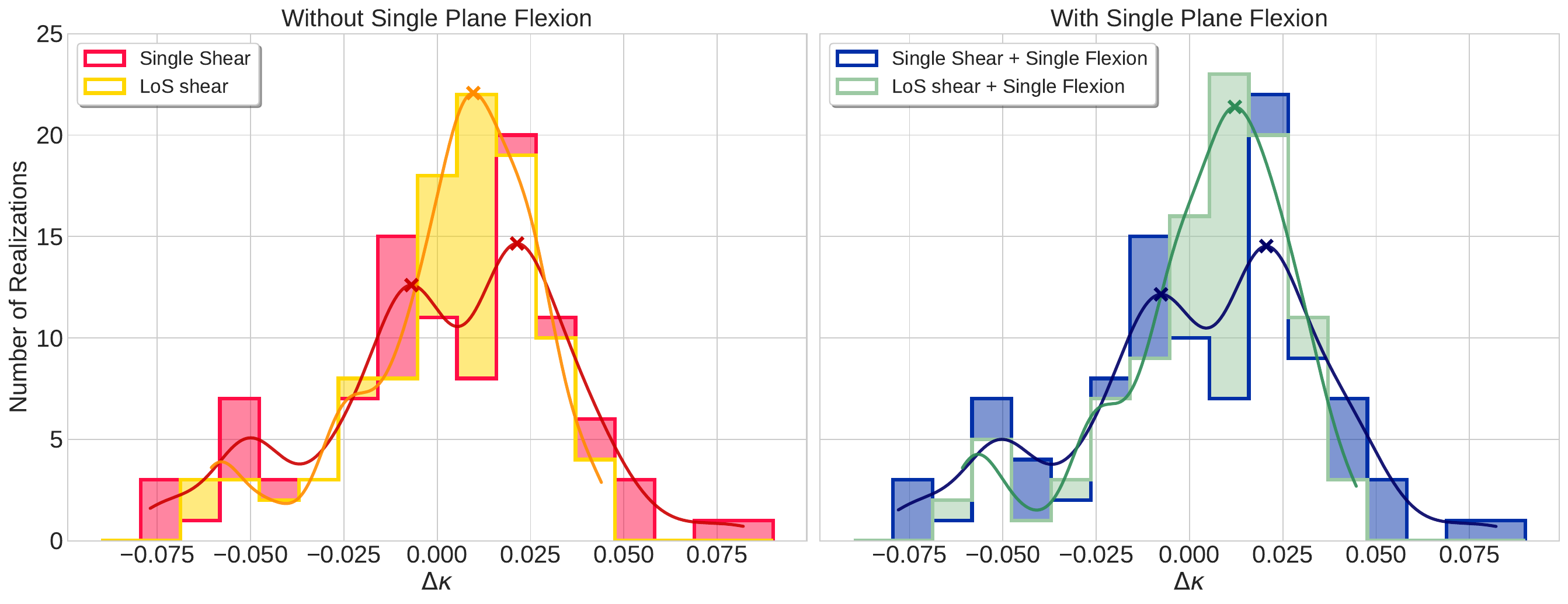}
\caption{Distributions of the convergence recovery residuals, $\Delta\kappa \equiv \kappa_{\mathrm{fit}} - \kappa_{\mathrm{simu}}$, across various mock realizations. The left panel shows the fitting results when single-plane flexion is excluded from the model, while the right panel displays the results with single-plane flexion included. In both panels, the distributions are compared for models assuming a single shear (red and blue) versus the LoS shear minimal model (yellow and green). The step histograms represent the absolute number of realizations, and the solid curves indicate the smoothed kernel density estimates, with crosses marking the peak of each distribution. Specifically, the single shear models exhibit bimodal distributions, peaking at approximately -0.007 and 0.021 for the Single Shear case, and at -0.008 and 0.021 for the Single Shear + Single Flexion case. Conversely, the LoS shear minimal models display unimodal distributions, peaking around 0.010 and 0.012 without and with Single Plane Flexion, respectively.
\label{fig:kappa_othetaE}}
\end{figure}

Figure~\ref{fig:kappa_othetaE} illustrates the distribution of the convergence modeling bias ($\Delta\kappa \equiv \kappa_{\rm fit} - \kappa_{\rm simu}$) across the mock realizations. Since $\theta_{\mathrm{E}}$ is fixed during the lens modeling, the recovered convergence parameter $\kappa_{\mathrm{fit}}$ effectively represents the ratio $\frac{1 - \kappa_{\mathrm{od}}}{1 - \kappa_{\mathrm{LoS}}}$, as derived in Appendix~\ref{sec:appendix_kinematics}. It should be noted that the definition of the true simulated convergence, $\kappa_{\rm simu}$, depends on the adopted modeling procedure. For the two single-plane models, $\kappa_{\rm simu}$ corresponds to the standard external convergence defined under the single-plane approximation, and was taken as $\kappa_{\mathrm{os}}$ in \cite{2021JCAP...08..024F}. In contrast, for the two multi-plane LoS models, it represents the total effective LoS convergence $\kappa_{\mathrm{LoS}}$, defined by Eq.~\eqref{eq: TidalLoS}. In all scenarios, these reference values are computed by averaging the underlying convergence within the typical strong lensing Einstein radius ($1.3 \pm 0.3 \, \mathrm{arcsec}$), derived using the ray-tracing method described in Section~\ref{subs:raytracing}.

In Fig.~\ref{fig:kappa_othetaE}, it is apparent that standard single-plane models (represented by the red and blue distributions) struggle to reproduce the true underlying convergence. Their error distributions are distinctly broader (with a median and 1$\sigma$ bias of $0.004^{+0.031}_{-0.030}$ without flexion, and $0.005^{+0.034}_{-0.031}$ with flexion) and exhibit a pronounced bimodal structure. As marked by the crosses on the smoothed kernel density estimates, the single-plane recovery consistently splits into two distinct peaks—one systematically underestimating the convergence at $\Delta\kappa \approx -0.007$, and the other overestimating it at $\Delta\kappa \approx 0.02$. Recall from Section~\ref{subsec:Foreground Shear} that freeing the lens ellipticity allows single-plane models to artificially absorb foreground shear perturbations to \textit{correct} the time-delay distance. However, Fig.~\ref{fig:kappa_othetaE} confirms that this mathematical compensation does not simultaneously guarantee an unbiased recovery of the convergence. The forced projection onto a single plane leaves residual systematic errors, which likely depend on the specific distribution of the line-of-sight perturbers.

In contrast, the explicit LoS shear minimal models, represented by the yellow and light green shaded regions, successfully mitigate these systematic errors. Whether modeling with or without single-plane flexion, the LoS approaches yield $\Delta\kappa$ biases that are significantly tighter (median and 1-$\sigma$ bias of $0.008^{+0.025}_{-0.022}$ and $0.008^{+0.026}_{-0.023}$, respectively). Furthermore, their distributions are characterized by a single, well-defined peak (located at $\Delta\kappa \approx 0.01$ for the base model and $\Delta\kappa \approx 0.012$ when flexion is included), effectively avoiding the bimodal structure seen in the single-plane cases. By accurately mapping the multi-plane shear field rather than approximating it as a single-plane perturbation, the LoS shear minimal modeling naturally breaks the geometric degeneracies that influence standard models. Consequently, explicitly accounting for LoS shear is essential for extracting the true external convergence, allowing for a robust correction of the MSD effect in strong lensing time-delay cosmography.

\subsection{The Role of Foreground Convergence}\label{subsec:Foreground convergence}

In our lens modeling of the mock realizations, the Einstein radius $\theta_{\rm E}$ is fixed. Nevertheless, as demonstrated by the LoS modeling results in the previous section, a residual bias of $\Delta\kappa \approx 0.008$ still persists in the recovered convergence. This remaining systematic error is primarily driven by the foreground convergence ($\kappa_{\mathrm{od}}$).

To understand why this bias persists despite both explicit LoS shear minimal modeling, we must examine the mathematical mismatch between stellar velocity dispersion measurements and the time-delay distance $D_{\rm dt}$ requirements~\citep{2022JCAP...07..027T, 2024JCAP...10..055J}. As detailed in Appendix~\ref{sec:appendix_kinematics}, while recovering the true $D_{\rm dt}$ requires correcting for the \textit{entire} mass sheet along the LoS, models with a fixed $\theta_{\rm E}$ are insensitive to the angular distortions introduced by foreground structures. Consequently, applying the total LoS correction to a model constrained by $\theta_{\rm E}$ leaves a residual systematic bias dictated by the foreground contribution:
\begin{equation}\label{eq:foregound}
    \frac{D_{\rm dt}^{\mathrm{fit}}}{D_{\rm dt}^{\mathrm{true}}} = 1 - \kappa_{\mathrm{od}}
\end{equation}

\begin{figure}
\centering
\includegraphics[width=1\textwidth]{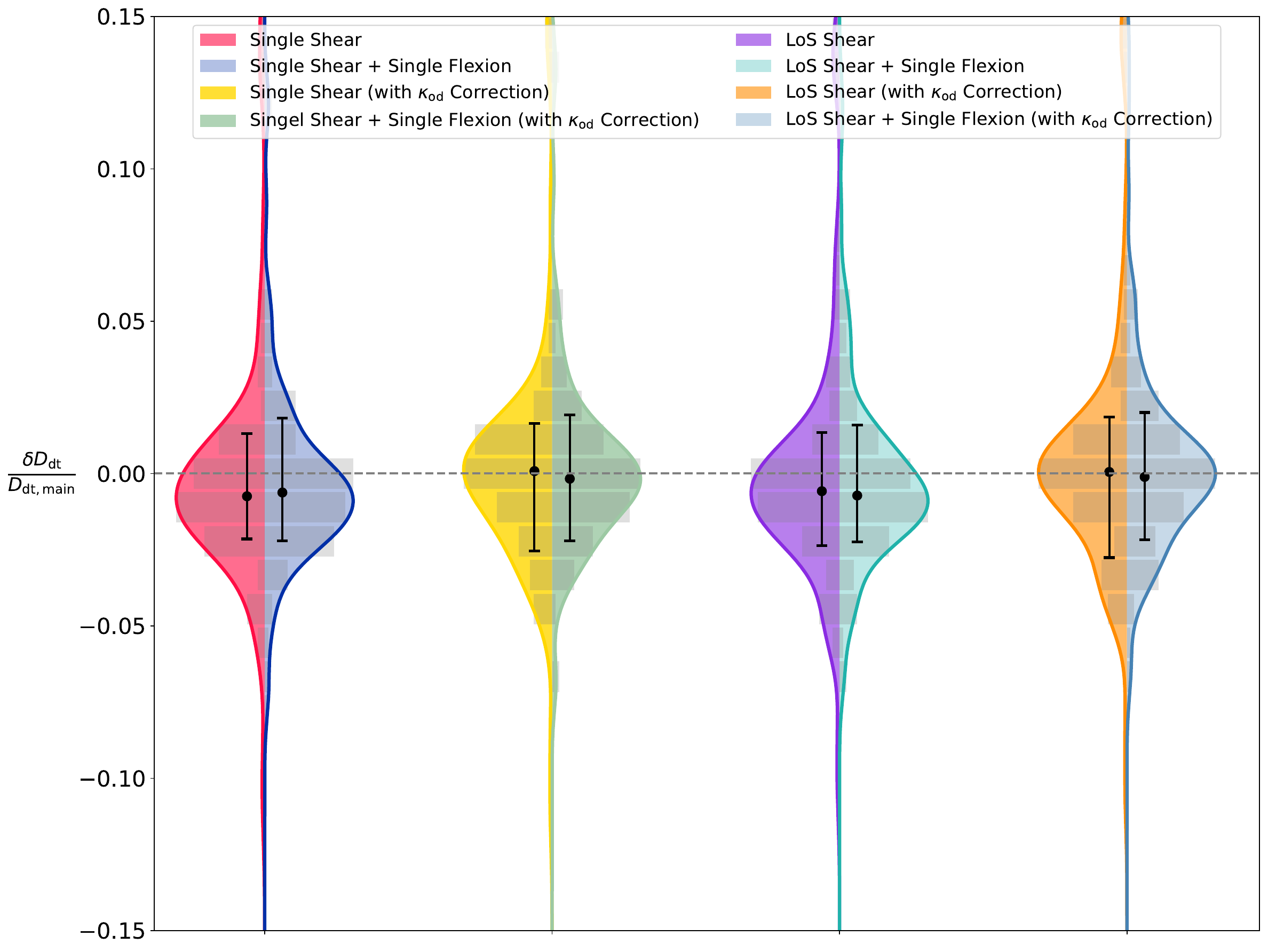}
\caption{Distribution of the modeling bias on the time-delay distance $D_{\rm dt}$ for 100 mock realizations, with only the Einstein radius $\theta_{\rm E}$ fixed. The plot displays four pairs of model configurations from left to right: uncorrected single-plane models (red/blue); single-plane models with $\kappa_{\rm od}$ correction (yellow/green); uncorrected LoS shear models (purple/teal); and LoS shear models with $\kappa_{\rm od}$ correction (orange/light blue). Other plot formatting details are similar to Fig.~\ref{fig:IntrinsicErrorAllFix}. Visually, both uncorrected model groups exhibit a systematic negative offset from the true value. Upon applying the $\kappa_{\rm od}$ correction, the distributions shift upwards and are precisely centered on zero. The detailed statistical analysis is presented in Table~\ref{table:stasbias}.
\label{fig:IntrinsicErrorFixThetaE}}
\end{figure}

To empirically validate this theoretical framework, Fig.~\ref{fig:IntrinsicErrorFixThetaE} presents the distribution of the fractional modeling bias on $D_{\rm dt}$ across our 100 mock realizations. Recall that the Einstein radius $\theta_{\rm E}$ is perfectly constrained in these fits, isolating the impact of the external mass sheet. The visual trend in Fig.~\ref{fig:IntrinsicErrorFixThetaE} aligns with our analytical expectations. Both the standard single-plane models (red and blue distributions) and the uncorrected explicit LoS shear models (purple and teal distributions) exhibit a clear, systematic underestimation of the true distance. This negative offset is the direct consequence of the missing foreground correction: as dictated by Equation \ref{eq:foregound}, a positive foreground convergence ($\kappa_{\rm od} > 0$), which is typical for these fields, inevitably leads to $D_{\rm dt}^{\mathrm{fit}} < D_{\rm dt}^{\mathrm{true}}$.

The visual evidence is strongly supported by the rigorous statistical analysis detailed in Table~\ref{table:stasbias}. Standard t-tests on the mean lack definitive significance here, due to the non-Gaussian nature of the distributions shown in Fig.~\ref{fig:IntrinsicErrorFixThetaE}. Consequently, we rely on the median and the Wilcoxon signed-rank test to robustly confirm the systematic bias. Further details on this statistical framework are deferred to Appendix~\ref{append:stat_tests}. For all uncorrected models, the median bias ranges from approximately $-0.6\%$ to $-0.8\%$. The Wilcoxon tests strongly reject the null hypothesis of zero bias in these cases, yielding $Z$-scores between $-2.20\sigma$ and $-2.52\sigma$ alongside $(1-p)$ values exceeding $0.97$. This confirms that the negative systematic offset is highly significant.

Crucially, once the explicit $\kappa_{\rm od}$ correction is applied, this systematic offset is successfully neutralized. As shown by the yellow/green and orange/light blue distributions in Fig.~\ref{fig:IntrinsicErrorFixThetaE}, the corrected distance estimates shift upwards and center precisely on the ground truth. Detailed in Table~\ref{table:stasbias}, the median biases for the corrected models shrink to negligible levels ($|\Delta| \lesssim 0.17\%$), and the corresponding Wilcoxon $(1-p)$ statistics drop dramatically to sub-$\sigma$ levels ($|Z| \le 0.65\sigma$), with $(1-p)$ values decreasing accordingly, indicating that the residual medians are statistically consistent with zero.
In summary, these results confirm that even when $\theta_{\rm E}$ is fixed to its true value and explicit LoS shear modeling is employed, time-delay cosmography remains fundamentally biased unless the foreground convergence is explicitly and accurately corrected. Although the magnitude of such biases currently remains subdominant within the broader TDCosmo uncertainty budget, implementing such corrections will be necessary for the high-precision era of next-generation surveys.

\section{conclusion and discussion}\label{Sec: Conclusion and discussion}

Time-delay cosmography offers a powerful, independent probe of the universe's expansion history. However, as the field drives toward percent-level precision to address cosmological crises such as the Hubble tension, mitigating systematic errors induced by LoS structures has become important. In this work, we evaluated the impact of multi-plane LoS perturbations on the recovery of the time-delay distance $D_{\rm dt}$ using a sample of 100 high-resolution mock quadruply-imaged realizations. We systematically compared the performance of standard single-plane external perturbation models against an explicit multi-plane LoS minimal lens model.

We find that, beyond the contribution of a constant convergence, large-scale matter clustering also induces surface mass density fluctuations within the SL Einstein radius.
We identify a critical region approximately 2 arcminutes in radius, within which the matter distribution determines the lensing potential inside the Einstein radius to better than 0.5\% accuracy.

By testing different model reconstruction strategies, our primary conclusions are summarized as follows:

\begin{itemize}
    \item We identified a severe degeneracy between foreground shear $ \gamma_{1,\rm od}, \gamma_{2,\rm od} $ and the main deflector's ellipticity $e_1, e_2$. When subjected to complex, multi-plane LoS perturbations, standard single-plane models completely fail to recover the true lens mass profile. While leaving the ellipticity parameters free allows these single-plane models to mathematically absorb the multi-plane effects and correct the time-delay distance, this comes at a catastrophic physical cost: the inferred lens shapes deviate from their true values by $10\sigma$ to $100\sigma$.
    \item Standard single-plane approximations fail to accurately recover the external convergence, resulting in excessively broad and distinctly bimodal error distributions for $\kappa$. Explicitly modeling the multi-plane perturbations via the LoS minimal lens model successfully breaks these geometric degeneracies. By properly accounting for foreground and background couplings, this approach yields tightly constrained reconstructions of both the main lens mass profile and the convergence from the LoS structures.
    \item Even when $\theta_{\rm E}$ is fixed to its true value and explicit LoS shear modeling is employed, time-delay distance estimates remain fundamentally biased if the foreground convergence ($\kappa_{\rm od}$) is neglected. Uncorrected models suffer from a systematic offset, underestimating the true distance by approximately $-0.6\%$ to $-0.8\%$, confirmed at a strong statistical significance of $2.2\sigma$ to $2.5\sigma$. However, explicitly applying the $\kappa_{\rm od}$ correction successfully neutralizes this systematic error, shrinking the residual median bias to statistically negligible, sub-$\sigma$ levels ($|\Delta| \lesssim 0.17\%$), and its statistical deviation from zero drops dramatically to strictly sub-sigma levels ($|Z| < 0.65\sigma$). This suggests that an accurate correction for foreground convergence is needed for future unbiased time-delay cosmography. While this sub-percent bias is currently subdominant compared to the overall uncertainty budget in contemporary TDCosmo analyses, our results emphasize that an accurate correction for foreground convergence will become necessary for future high-precision, unbiased time-delay cosmography.
\end{itemize}

\begin{figure}
\centering
\includegraphics[width=1\textwidth]{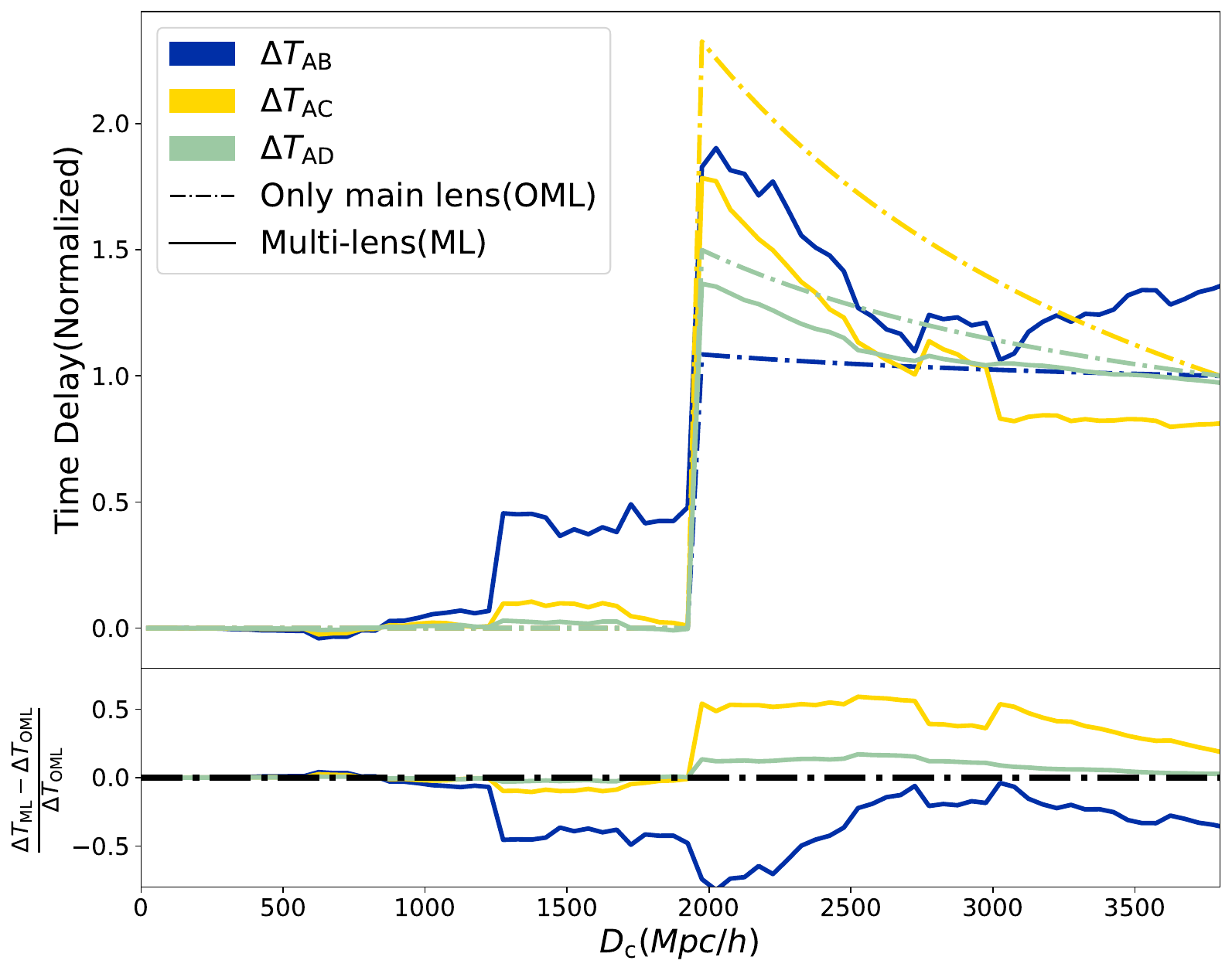}
\caption{Ray-traced cumulative time delays from the observer to the multiply imaged source, normalized by the time delay in the source plane with only main lens. The top panel shows the evolution of the normalized time delays for three image pairs: $\Delta T_{\mathrm{AB}}$ (blue), $\Delta T_{\mathrm{AC}}$ (yellow), and $\Delta T_{\mathrm{AD}}$ (green) as a function of comoving distance $D_c$, calculated using Eqs.~\eqref{eq:timedelay1} and \eqref{eq:timedelay2}. Solid lines represent the results from the multi-lens (ML) simulation, which incorporates line-of-sight structures, while dash-dotted lines indicate the idealized case with only the main lens (OML). The bottom panel displays the relative fractional difference between the two scenarios, defined as $(\Delta T_{\mathrm{ML}} - \Delta T_{\mathrm{OML}}) / \Delta T_{\mathrm{OML}}$, highlighting the cumulative perturbations introduced by line-of-sight multi-plane lensing effects.
\label{fig:CDF of TD}}
\end{figure}

To better understand the findings above, we present the cumulative time delays back in time at each lens plane for a typical system in Fig. \ref{fig:CDF of TD}, calculated using Eqs.~\eqref{eq:timedelay1} and \eqref{eq:timedelay2}. To facilitate comparison between different image pairs, the cumulative time delays for each pair are normalized by their respective total time delays in the single lens approximation.
In this specific sample, the dash-dotted lines represent the case where only the main lens is considered. Different colors correspond to time delays between different image pairs. In the single-lens scenario, a sharp rise in the time delay occurs at the main lens plane, primarily due to the lensing potential term. 
The smoothed decline reflects the fact that, before reaching the main lens plane, the light rays of different images propagate in different directions toward the observer, leading to the accumulation of geometric time delay. After passing through the main lens plane, the light rays propagate in the constant directions toward the observer. Hence, the differential time delay vanishes according to Eq.~\eqref{eq:timedelay1}. In contrast, the solid curves show the cumulative time delays from the observer to the source, incorporating multiple lens planes. The inclusion of multiple planes introduces two key differences. First, the lensing potential terms change because the additional planes perturb the light rays’ positions at the main lens. Second, both the deflection and potential of the intermediate planes contribute to the cumulative time delays. As illustrated in the lower panel of Fig. \ref{fig:CDF of TD}, the differential time delays at each lens plane can deviate by as much as $50\%$ compared to the single-lens scenario. Furthermore, since we normalize the time delay of each image pair using the total time delay from the single-lens approximation, the cumulative time delay in the multiple-lens case does not necessarily match that of the single-lens model. This discrepancy is reflected in the fact that the cumulative time delay curve does not converge to unity.

These findings have significant implications for the future of strong lensing cosmography. The bimodal error distributions in the recovered convergence highlight a fundamental limitation of treating large-scale structure as a simple, single-plane phase sheet. Because the fitted time-delay distance scales inversely with $(1 - \kappa)$, any systematic bias or artificially broadened uncertainty in $\kappa$ directly propagates into the final cosmological constraints, severely limiting our ability to correct for the MSD.
These findings have significant implications for the future of strong lensing cosmography. The bimodal error distributions in the recovered convergence highlight a fundamental limitation of treating large-scale structure as a simple, single-plane phase sheet. Because the fitted time-delay distance scales inversely with $(1 - \kappa)$, any systematic bias or artificially broadened uncertainty in $\kappa$ directly propagates into the final cosmological constraints, severely limiting our ability to correct for the MSD.

In this work, we fixed the macro-model parameter $\theta_{\mathrm{E}}$ to isolate the MSD effect, which introduces a bias stemming from foreground convergence. When comparing this approach with kinematic measurements, where the velocity dispersion is modeled as~\citep{2020A&A...639A.101M,2016JCAP...08..020B,2019MNRAS.484.4726B}
\begin{equation}
    \sigma^2(\boldsymbol{\theta}) \propto \left( \frac{1 - \kappa_{\mathrm{LOS}}}{1 - \kappa_{\mathrm{od}}} \right) \frac{D_{\mathrm{os}}}{D_{\mathrm{ds}}} J(\theta, \theta_{\mathrm{E}}, \gamma)\;.
\end{equation}
By comparing the measured velocity dispersion $\sigma^2(\boldsymbol{\theta})$ with the theoretical prediction $(D_{\mathrm{os}}/D_{\mathrm{ds}}) J(\theta, \theta_{\mathrm{E}}, \gamma)$, one obtains the convergence ratio $(1 - \kappa_{\mathrm{LOS}})/(1 - \kappa_{\mathrm{od}})$. This result is conceptually equivalent to our approach of fixing $\theta_{\mathrm{E}}$. Notably, while the angular diameter distance ratio $D_{\mathrm{os}}/D_{\mathrm{ds}}$ depends weakly on certain cosmological parameters, such as the matter density $\Omega_m$, it remains entirely independent of the Hubble constant, which is our major concern. Thus, by incorporating kinematic measurements, one can place a direct constraint on the ratio $(1 - \kappa_{\mathrm{LOS}})/(1 - \kappa_{\mathrm{od}})$, which biases the time-delay modelling if not properly accounted.

While this study establishes the critical necessity of explicit LoS modeling, several avenues remain for future exploration. In our current framework, we simplified the treatment of higher-order multi-plane effects by adopting a single-plane approximation for flexion to avoid over-parameterization. Interestingly, we found that incorporating this single-plane flexion did not yield a noticeable improvement in tightening the constraints on $D_{\rm dt}$). This indicates that such a simplified treatment might be insufficient to fully capture complex LoS perturbations or higher order fluctuations. Therefore, future work would investigate whether incorporating explicit multi-plane flexion terms~\citep{2024JCAP...08..021D}, or non-parametric methods to capture the fluctuations shown in Fig.~\ref{fig:case_los}, could further tighten the constraints on highly perturbed, asymmetric systems.

\appendix

\section{Derivation of the Effective LoS Potential} 
\label{app:LoS_potential}

To evaluate the sensitivity to the relevant angular scales, we derive the effective LoS lensing potential used in Eq.~\eqref{eq:eff_potential}. By neglecting higher-order lens-lens couplings and isolating the LoS contributions, the effective potential associated with the multi-plane displacement field is:
\begin{equation}
    \psi(\boldsymbol{\theta}) \approx  \sum_{i \ne {\rm d}} \frac{D_{i\mathrm{s}}}{D_{\mathrm{s}} D_i} \hat{\psi}_i(D_i \boldsymbol{\theta}_i),
\end{equation}
where $\hat{\psi}_i$ is the projected Newtonian potential of the $i$-th lens plane. 

Our dimensionless lensing potential $\psi_i$ is defined via the Poisson equation $\nabla^2 \psi_i = 2\kappa_i$, where $\kappa_i$ is from Eq.~\eqref{eq:kappa_slice} and assumes a source at infinity ($s \rightarrow \infty$). Thus, it relates to $\hat{\psi}_i$ by taking the infinite source limit:
\begin{equation}
    \psi_i = \lim_{s \rightarrow \infty} \frac{D_{i\mathrm{s}}}{D_i D_{\mathrm{s}}} \hat{\psi}_i = \frac{1}{D_i} \hat{\psi}_i.
\end{equation}

Substituting $\hat{\psi}_i = D_i \psi_i$ back into the total potential, the $D_i$ factors cancel out:
\begin{equation}
    \psi(\boldsymbol{\theta}) =  \sum_{i \ne {\rm d}} \frac{D_{i\mathrm{s}}}{D_{\mathrm{s}}} \psi_i(\boldsymbol{\theta}_i).
\end{equation}


\section{The fluctuations in convergence and shear}\label{append:std}

\begin{figure}[ht!]
\centering
\includegraphics[width=1\textwidth]{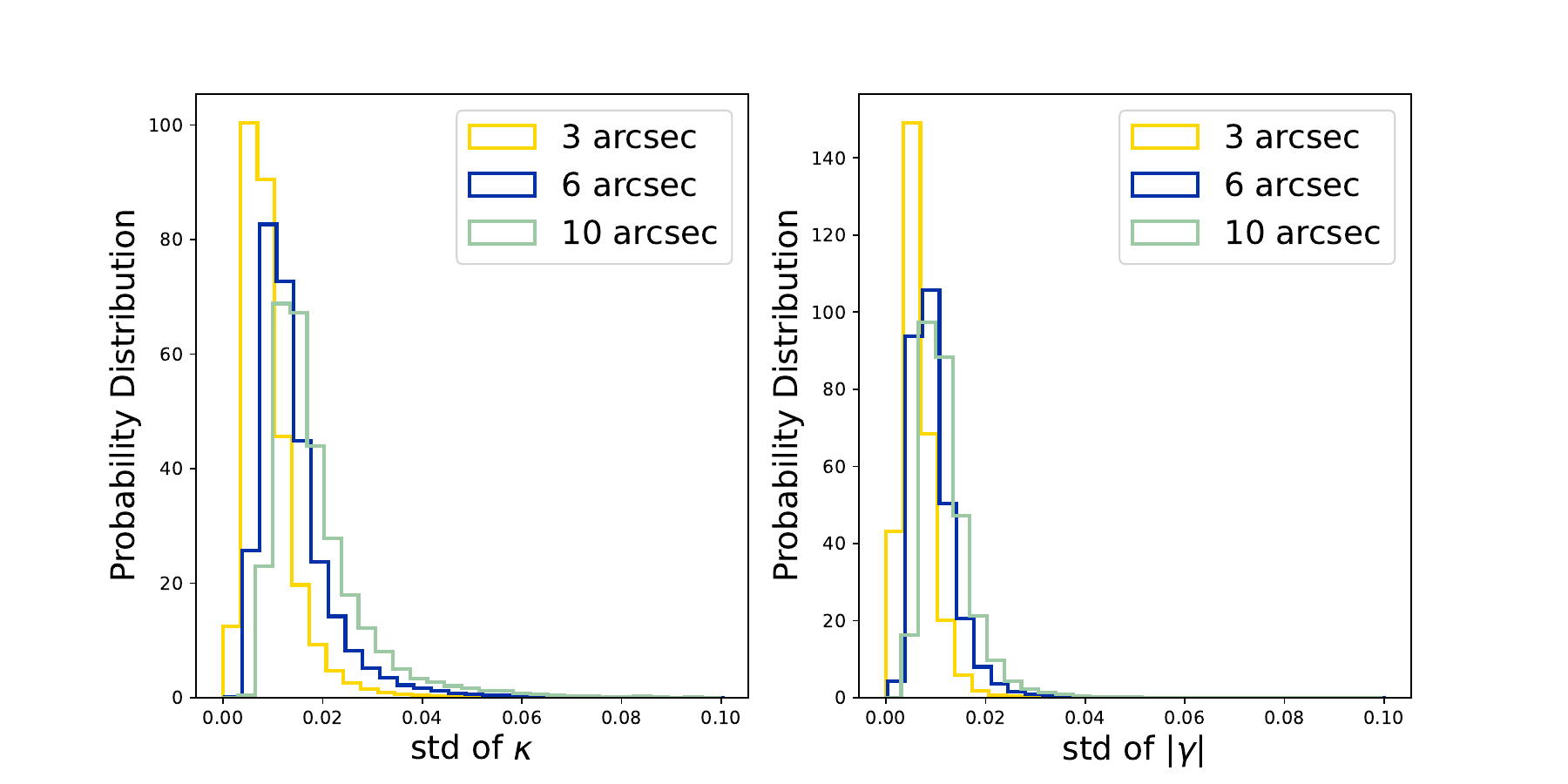}
\caption{Probability distributions of the standard deviation of the LoS convergence (left panel) and shear magnitude (right panel). The yellow, blue, and green histograms correspond to measurements evaluated within angular apertures of 3, 6, and 10 arcseconds, respectively.}
\label{fig:std of LoS}
\end{figure}

We quantify the spatial fluctuations of the LoS lensing quantities at different angular scales. Figure~\ref{fig:std of LoS} presents the probability distributions of the standard deviation of the convergence in the left panel and the magnitude of the shear in the right panel, evaluated within varying angular apertures (3, 6, and 10 arcseconds).

\section{The Foreground Convergence}
\label{sec:appendix_kinematics}

For a lens with a generic power-law density profile $\rho \propto r^{-\gamma}$, the observable geometric Einstein radius $\theta_{\rm E}^{\rm obs}$ is scaled by both the foreground convergence $\kappa_{\mathrm{od}}$ and the total line-of-sight convergence $\kappa_{\mathrm{LoS}}$. It relates to the true Einstein radius $\theta_{\rm E}^{\rm true}$ via:
\begin{equation}
\theta_{\rm E}^{\rm obs} = \left[ \frac{(1 - \kappa_{\mathrm{od}})^{3-\gamma}}{1 - \kappa_{\mathrm{LoS}}} \right]^{\frac{1}{\gamma - 1}} \theta_{\rm E}^{\rm true}.
\end{equation}
Specifically, for the SIE profile ($\gamma = 2$) utilized in our mock analysis, this geometric scaling elegantly simplifies to:
\begin{equation}
\theta_{\rm E}^{\rm obs} = \theta_{\rm E}^{\rm true} \left( \frac{1 - \kappa_{\mathrm{od}}}{1 - \kappa_{\mathrm{LoS}}} \right).
\end{equation}

Conversely, the true time-delay distance $D_{\rm dt}^{\mathrm{true}}$ (where $H_0 \propto 1/D_{\rm dt}$) strictly requires a correction for the \textit{entire} mass sheet along the line of sight:
\begin{equation}\label{eq:ddtoftrue}
    D_{\rm dt}^{\mathrm{true}} = \frac{D_{\rm dt}^{\mathrm{model}}}{1 - \kappa_{\mathrm{LoS}}}
\end{equation}
where $D_{\rm dt}^{\mathrm{model}}$ is the distance inferred from a lens model neglecting external convergence.

By fixing the Einstein radius to its true value, $\theta_{\rm E} = \theta_{\rm E}^{\rm true}$, the inferred time-delay distance becomes:
\begin{equation}
    D_{\rm dt}^{\mathrm{fit}} = D_{\rm dt}^{\mathrm{model}} \left( \frac{1 - \kappa_{\mathrm{od}}}{1 - \kappa_{\mathrm{LoS}}} \right)
\end{equation}
Comparing this inferred distance to the true requirement Eq.~\eqref{eq:ddtoftrue} yields the residual systematic bias presented in Section \ref{subsec:Foreground convergence}:
\begin{equation}
    \frac{D_{\rm dt}^{\mathrm{fit}}}{D_{\rm dt}^{\mathrm{true}}} = 1 - \kappa_{\mathrm{od}}
\end{equation}
For details, we prefer reader to~\citep{2024JCAP...10..055J}.


\section{Statistical tests for sample mean}\label{append:stat_tests}
To evaluate whether a sample exhibits a significant deviation of its central tendency from a reference value (typically zero), we consider two commonly used statistical tests: the one-sample t-test and the Wilcoxon signed-rank test. These tests differ in their assumptions about the underlying distribution.

The one-sample t-test is appropriate when the sample values \(X = \{x_1, x_2, \dots, x_n\}\) are approximately normally distributed. The test examines the null hypothesis
\begin{equation}
H_0: \mu = \mu_0,
\end{equation}
where \(\mu\) is the population mean and \(\mu_0\) is the hypothesized reference. The test statistic is defined as
\begin{equation}
t = \frac{\bar{x} - \mu_0}{s / \sqrt{n}},
\end{equation}
where \(\bar{x} = \frac{1}{n} \sum_{i=1}^{n} x_i\) is the sample mean, \(s = \sqrt{\frac{1}{n-1} \sum_{i=1}^{n} (x_i - \bar{x})^2}\) is the sample standard deviation, and \(n\) is the sample size\footnote{We use $n-1$ in the denominator, applying Bessel's correction to compute the unbiased sample variance.}. Under the null hypothesis, \(t\) follows a Student's t-distribution with \(n-1\) degrees of freedom. The resulting p-value quantifies the probability of observing a test statistic at least as extreme as the calculated $t$, assuming $H_0$ is true.

For samples that deviate from normality or exhibit skewness, the Wilcoxon signed-rank test provides a non-parametric alternative. To compute the \textit{W-statistic} (reported as \texttt{W-stat}), we calculate the differences from the theoretical zero-bias reference, $d_i = x_i - \mu_0$, discarding any zero differences. The absolute differences $|d_i|$ are then ranked in ascending order to assign ranks $R_i$. The W-statistic is defined as the sum of the ranks of the positive differences, $W = \sum_{d_i>0} R_i$.

For large sample sizes ($n \gtrsim 20$), the distribution of the W-statistic is well-approximated by a normal distribution with the z-score ($Z$), which follows a standard normal distribution, $\mathcal{N}(0,1)$:
\begin{equation}
    Z = \frac{W - \frac{n(n+1)}{4}}{\sqrt{\frac{n(n+1)(2n+1)}{24}}}
\end{equation}
The $1-p$ value is then derived from $Z$.
\label{sec:appendix_stats}

\vspace{2mm}
\noindent
{\textbf{Data availability}}
The \texttt{ELUCID} N-body simulation is public at \url{https://www.elucid-project.com}.
The mock images and time delays with LoS effects are publicly available via Zenodo at \url{https://doi.org/10.5281/zenodo.20512184}. 

\vspace{2mm}
\noindent
{\textbf{Code availability}}
The mock data generation code is released under the MIT license and archived on Zenodo at \url{https://doi.org/10.5281/zenodo.20512184}. The model reconstruction was performed using \texttt{lenstronomy} (\url{https://lenstronomy.readthedocs.io}).

\vspace{2mm}
\noindent
{\textbf{Acknowledgements}}
We are grateful to Nan Li, Tian Li, and Xikai Shan for insightful discussions. We also thank Yifan Chen for creating the illustrative cartoon in Fig.~\ref{fig:RaysCartoon}.
Furthermore, we thank the referee for their constructive comments and suggestions, which significantly improved the quality of this manuscript.
This work is supported in part by the National Natural Science Foundation of China Grants No. 12333001 and No. 11903082.  
We acknowledge the cosmology simulation database (CSD) in the National Basic Science Data Center (NBSDC) and its funds the NBSDC-DB-10.

\vspace{2mm}
\noindent
{\textbf{Author contributions}}
All authors provided ideas throughout the project and comments on the manuscript.
SL contributed in calculating and writing the draft.
BH contributed in proposing the idea and writing the draft.
CW contributed in generating the lensing catalogs.

\vspace{2mm}
\noindent
{\textbf{Competing interests}}
The authors have no competing interests.

\bibliography{sample631}{}
\bibliographystyle{aasjournal}


\end{document}